\documentclass[nojss]{jss}

\usepackage[utf8]{inputenc}

\providecommand{\tightlist}{%
  \setlength{\itemsep}{0pt}\setlength{\parskip}{0pt}}

\author{
Nicole S. Erler\\Erasmus Medical Center \And Dimitris Rizopoulos\\Erasmus Medical Center \And Emmanuel M.E.H. Lesaffre\\KU Leuven
}
\title{\pkg{JointAI}: Joint Analysis and Imputation of Incomplete Data in
\proglang{R}}

\Plainauthor{Nicole S. Erler, Dimitris Rizopoulos, Emmanuel M.E.H. Lesaffre}
\Plaintitle{JointAI: Joint Analysis and Imputation of Incomplete Data in R}
\Shorttitle{\pkg{JointAI}: Joint Analysis and Imputation}

\Abstract{
Missing data occur in many types of studies and typically complicate the
analysis. Multiple imputation, either using joint modelling or the more
flexible fully conditional specification approach, are popular and work
well in standard settings. In settings involving non-linear associations
or interactions, however, incompatibility of the imputation model with
the analysis model is an issue often resulting in bias. Similarly,
complex outcomes such as longitudinal or survival outcomes cannot be
adequately handled by standard implementations. In this paper, we
introduce the \proglang{R} package \pkg{JointAI}, which utilizes the
Bayesian framework to perform simultaneous analysis and imputation in
regression models with incomplete covariates. Using a fully Bayesian
joint modelling approach it overcomes the issue of uncongeniality while
retaining the attractive flexibility of fully conditional specification
multiple imputation by specifying the joint distribution of analysis and
imputation models as a sequence of univariate models that can be adapted
to the type of variable. \pkg{JointAI} provides functions for Bayesian
inference with generalized linear and generalized linear mixed models
and extensions thereof as well as survival models and joint models for
longitudinal and survival data, that take arguments analogous to
corresponding well known functions for the analysis of complete data
from base \proglang{R} and other packages. Usage and features of
\pkg{JointAI} are described and illustrated using various examples and
the theoretical background is outlined.
}

\Plainkeywords{imputation, Bayesian, missing covariate, non-linear, interaction, multi-level, survival, joint model, R, JAGS}

\Address{
    Nicole S. Erler\\
  Erasmus Medical Center\\
  Department of Biostatistics\newline Doctor Molewaterplein
  40\newline 3015 GD Rotterdam, the Netherlands\\
  E-mail: \email{n.erler@erasmusmc.nl}\\
  URL: www.nerler.com\\~\\
      }

\usepackage{amsmath} \usepackage{booktabs} \usepackage{dsfont} \usepackage{placeins} \usepackage[T1]{fontenc}

\begin{document}

\hypertarget{introduction}{%
\section{Introduction}\label{introduction}}

Missing data are a challenge common to the analysis of data from
virtually all kinds of studies. Especially when many variables are
measured, as in large cohort studies, or when data are obtained
retrospectively, e.g., from registries, large proportions of missing
values in some variables are not uncommon.

Multiple imputation, which appears to be the gold standard to handle
incomplete data, as indicated by its widespread use, has its origin in
the 1970s and was primarily developed for survey data
\citep{Deng2016, Treiman2009,
Rubin1987, Rubin2004}. One of its first implementations in \proglang{R}
\citep{RVersion} is the package \pkg{norm} \citep{norm}, which performs
multiple imputation under the joint modelling framework using a
multivariate normal distribution \citep{Schafer1997}. Nowadays more
frequently used is multiple imputation using a fully conditional
specification (FCS), also called multiple imputation using chained
equations (MICE) and its seminal implementation in the \proglang{R}
package \pkg{mice} \citep{mice, Buuren2012}.

Since the introduction of multiple imputation, datasets have gotten more
complex. Therefore, more sophisticated methods that can adequately
handle the features of modern data and comply with assumptions made in
its analysis are required. Modern studies do not only record univariate
outcomes, measured in a cross-sectional setting but also outcomes that
consist of two or more measurements, for instance, repeatedly measured
or survival outcomes. Furthermore, non-linear effects, introduced by
functions of covariates, such as transformations, polynomials or
splines, or interactions between variables are considered in the
analysis and, hence, need to be taken into account during imputation.

Standard multiple imputation, either using FCS or a joint modelling
approach, e.g., under a multivariate normal distribution, assumes linear
associations between all variables It is possible to include non-linear
associations using transformations of variables and passive imputation
\citep{Buuren2012}; however, this does not generally solve the issue of
uncongenial and/or incompatible imputation models. Moreover, FCS
requires the outcome to be explicitly specified in each of the linear
predictors of the full conditional distributions. In settings where the
outcome is more complex than just univariate, for instance, for a
survival outcome that typically is represented by the observed event or
censoring time and a censoring indicator, or a longitudinal outcome
consisting of multiple, correlated measurements, this is not
straightforward and not generally possible without information loss,
leading to misspecified imputation models and, likely, to bias.

Some extensions of standard multiple imputation have been developed and
are implemented in \proglang{R} packages and other software, but the
greater part of the software for imputation is restricted to standard
settings such as cross-sectional survey data. The Comprehensive
\proglang{R} Archive Network (CRAN) task view on missing data
(\url{https://CRAN.R-project.org/view=MissingData}) gives an overview of
available \proglang{R} packages that deal with missing data in different
contexts, using various approaches.

Relevant in our context, i.e., in settings where potentially complex
models, such as models with non-linear associations, survival outcomes
or multi-level structure, are estimated on data with missing values in
covariates, are for example the following \proglang{R} packages.

The package \pkg{mice} itself provides limited options to perform
multi-level imputation, restricted to conditionally normal and binary
level-1 covariates (e.g., repeated measurements) and the use of a linear
model or predictive mean matching for level-2 covariates (e.g.,
patient-specific characteristics). The packages \pkg{micemd}
\citep{micemd} and \pkg{miceadds} \citep{miceadds} provide extensions to
Poisson models and predictive mean matching for level-1 covariates.

\pkg{smcfcs} \citep{smcfcs}, short for ``substantive model compatible
fully conditional specification'', uses Bayesian methodology to extend
standard multiple imputation using FCS to ensure compatibility between
analysis model and imputation models. It can handle linear, logistic and
Poisson models, as well as parametric (Weibull) and Cox proportional
hazards survival models, and competing risk models. Additionally, it
provides functionality for case cohort and nested case control studies.
The model specification is similar to the \pkg{mice} package, however
less automated.

The \proglang{R} package \pkg{jomo} \citep{jomo} performs joint model
multiple imputation in the Bayesian framework using a multivariate
normal distribution and includes an extension to the standard approach
to assure compatibility between analysis model and imputation models. It
can handle generalized linear (mixed) models, cumulative link mixed
models, proportional odds probit regression and Cox proportional hazards
models. Unfortunately, no functions are available to facilitate the
evaluation of convergence of the Markov chain Monte Carlo (MCMC)
algorithm. The \proglang{R} package \pkg{mitml} \citep{mitml} provides
an interface to \pkg{pan} (imputation of continuous level-1 covariates
only) and \pkg{jomo} and includes functions that make the analysis and
evaluation of the imputed data more convenient.

\pkg{hmi} \citep{hmi} (``hierarchical multi-level imputation'') combines
functionality of the packages \pkg{mice} and \pkg{MCMCglmm}
\citep{MCMCglmm} to perform multiple imputation in single- and
multi-level models, but it assumes all incomplete covariates in
multi-level models to be level-1 covariates. Similarly, the package
\pkg{mlmmm} \citep{mlmmm}, which uses the EM-algorithm to perform
multi-level imputation, does not consider incomplete level-2 variables.

\pkg{mdmb} \citep{mdmb} implements model-based treatment of missing data
using likelihood or Bayesian methods in linear and logistic regression
and linear and ordinal multi-level models. Under the Bayesian framework,
substantive model compatible imputation is available. A drawback is that
the specification does not follow the specification of well-known
\proglang{R} functions, which complicates usage especially for new
users, and that the specification of more complex models can quickly
become quite involved.

Depending on the type of outcome model (survival, multi-level or
single-level), whether non-linear effects are involved (which need
substantive model compatible imputation), the measurement level of
incomplete covariates and whether missingness occurs in level-1 (e.g.,
repeated measurements) as well as in level-2 covariates (e.g., baseline
covariates), the user has to work with different software packages. This
requires users to be familiar with the usage and underlying statistical
methodology of a number of packages and approaches. Since for several
packages the documentation is rather inscrutable and vague, it is
unclear what precisely these packages can and cannot do and what the
underlying assumptions are. Choosing an appropriate software package and
applying it correctly may, thus, become quite a daunting challenge.

The \proglang{R} package \pkg{JointAI} \citep{JointAI}, which is
presented in this paper, aims to facilitate the correct analysis of
incomplete data by providing a unified framework for both simple and
more complex models, using a consistent specification that most users
will be familiar with from commonly used (base) \proglang{R} functions.

Most of the packages named above perform multiple imputation, i.e.,
create multiple imputed datasets, which are then analysed in a second
step, followed by pooling of the results. While the separation of
imputation and analysis is often considered an advantage, especially
when large databases are to be analysed by multiple researchers, this
separation permits the use of analysis models that are incompatible with
the imputation models. \pkg{JointAI} follows a different, fully Bayesian
approach (used as well in \pkg{mdmb}). By modelling the analysis model
of interest jointly with the incomplete covariates, analysis and
imputation can be performed simultaneously while assuring compatibility
between all sub-models \citep{Erler2016, Erler2019}. In this joint
modelling approach, the added uncertainty due to the missing values is
automatically taken into account in the posterior distribution of the
parameters of interest, and no pooling of results from repeated analyses
is necessary. The joint distribution is specified conveniently, using a
sequence of conditional distributions that can be specified flexibly
according to each type of variable. Since the analysis model of interest
defines the first distribution in the sequence, the outcome is included
in the joint distribution without the need for it to enter the linear
predictor of any of the other models. Moreover, non-linear associations
that are part of the analysis model are automatically taken into account
for the imputation of missing values. This directly enables our approach
to handle complicated models, with complex outcomes and flexible linear
predictors. Another feature that distinguishes \pkg{JointAI} from the
other packages named above is that it can handle hierarchical settings
with more than two levels.

In this paper, we introduce the \proglang{R} package \pkg{JointAI},
which performs joint analysis and imputation of regression models with
incomplete covariates under the missing at random (MAR) assumption
\citep{Rubin1976}, and explain how data with incomplete covariate
information can be analysed and imputed with it. The package is
available for download at the Comprehensive \proglang{R} Archive Network
(CRAN) under \url{https://CRAN.R-project.org/package=JointAI}. Section 2
briefly describes the theoretical background of the method. An outline
of the general structure of \pkg{JointAI} is given in Section 3,
followed by an introduction of the example datasets that are used
throughout the paper in Section 4. Details about model specification,
settings controlling the MCMC sampling, and summary, plotting and other
functions that can be applied after fitting the model are given in
Sections 5 through 7. We conclude the paper with an outlook of planned
extensions and discuss the limitations that are introduced by the
assumptions made in the fully Bayesian approach.

\hypertarget{theoretical-background}{%
\section{Theoretical background}\label{theoretical-background}}

Consider the general setting of a regression model where interest lies
in a set of parameters \(\boldsymbol\theta\) that describe the
association between a univariate outcome \(\mathbf y\) and a set of
covariates \(\mathbf X = (\mathbf x_1, \ldots, \mathbf x_p)\). In the
Bayesian framework, inference over \(\boldsymbol\theta\) is obtained by
estimation of the posterior distribution of \(\boldsymbol\theta\), which
is proportional to the product of the likelihood of the data
\((\mathbf y, \mathbf X)\) and the prior distribution of
\(\boldsymbol\theta\),
\[ p(\boldsymbol\theta\mid \mathbf y, \mathbf X) \propto
p(\mathbf y, \mathbf X \mid \boldsymbol\theta)\,p(\boldsymbol\theta).\]

When some of the covariates are incomplete, \(\mathbf X\) consists of
two parts, the completely observed variables \(\mathbf X_{obs}\) and
those variables that are incomplete, \(\mathbf X_{mis}\). If
\(\mathbf y\) had missing values (and this missingness was ignorable),
the only necessary change in the formulas below would be to write
\(\mathbf y_{mis}\) instead of \(\mathbf y\), however the model itself
would not change, since the conditional distribution for \(y\) is
already part of the model specification. Here, we will, therefore,
consider \(\mathbf y\) to be completely observed. In the implementation
in the \proglang{R} package \pkg{JointAI}, however, missing values in
the outcome are allowed and are imputed automatically.

The likelihood of the complete data, i.e., observed and unobserved data,
can be factorized in the following convenient way:
\[p(\mathbf y, \mathbf X_{obs}, \mathbf X_{mis} \mid \boldsymbol\theta) =
p(\mathbf y \mid \mathbf X_{obs}, \mathbf X_{mis}, \boldsymbol\theta_{y\mid x})\,
p(\mathbf X_{mis} \mid \mathbf X_{obs}, \boldsymbol\theta_x),
\] where the first factor constitutes the analysis model of interest,
described by a vector of parameters \(\boldsymbol\theta_{y\mid x}\), and
the second factor is the joint distribution of the incomplete variables,
i.e., the imputation part of the model, described by parameters
\(\boldsymbol\theta_x\), and
\(\boldsymbol\theta = (\boldsymbol\theta_{y\mid x}^\top, \boldsymbol\theta_x^\top)^\top\).

Explicitly specifying the joint distribution of all data is one of the
major advantages of the Bayesian approach, since this facilitates the
use of all available information of the outcome in the imputation of the
incomplete covariates \citep{Erler2016}, which becomes especially
relevant for more complex outcomes like repeatedly measured variables
(see Section~\ref{sec:impLong}).

In complex models the posterior distribution can usually not be derived
analytically but MCMC methods are used to obtain samples from the
posterior distribution. The MCMC sampling in \pkg{JointAI} is done using
the Gibbs method, which iteratively samples from the full conditional
distributions of the unknown parameters and missing values.

In the following sections we describe each of the three parts of the
model, the analysis model, the imputation part and the prior
distributions, in detail.

\hypertarget{sec:AnalysisModel}{%
\subsection{Analysis model}\label{sec:AnalysisModel}}

The analysis model of interest is described by the probability density
function \(p(\mathbf y \mid \mathbf X, \boldsymbol\theta_{y\mid x})\).
The \proglang{R} package \pkg{JointAI} can currently handle analysis
models that are generalized linear regression models (GLM) or
generalized linear mixed models (GLMM) or extensions thereof (using
either a log-normal or a beta distribution), cumulative and multinomial
logit (mixed) models, parametric (Weibull) or proportional hazards
survival models. Moreover, it is possible to fit joint models for
longitudinal and survival data.

In a multi-level setting, we use level-1 to refer to the lowest level of
the hierarchy, for instance, repeated measurements of a biomarker,
level-2 to the next higher level (e.g., patient-specific information),
and so on. \pkg{JointAI} allows for models with more than two levels,
but, to facilitate notation, we focus here on settings with two levels.

\hypertarget{generalized-linear-mixed-models}{%
\subsubsection{Generalized linear (mixed)
models}\label{generalized-linear-mixed-models}}

For a GLM the probability density function is chosen from the
exponential family and has the linear predictor
\[g\{E(y_i\mid \mathbf X, \boldsymbol\theta_{y\mid x})\} = \mathbf x_i^\top\boldsymbol\beta,\]
where \(g(\cdot)\) is a link function, \(y_i\) the value of the outcome
variable for subject \(i\), and \(\mathbf x_i\) is a column vector
containing the row of \(\mathbf X\) that contains the covariate
information for \(i\).

For a GLMM the linear predictor is of the form
\[g\{E(y_{ij}\mid \mathbf X,
\mathbf b_i, \boldsymbol\theta_{y\mid x})\} = \mathbf
x_{ij}^\top\boldsymbol\beta + \mathbf z_{ij}^\top\mathbf b_i,\] where
\(y_{ij}\) is the \(j\)-th outcome of subject \(i\), \(\mathbf x_{ij}\)
is the corresponding vector of covariate values, \(\mathbf b_i\) a
vector of random effects pertaining to subject \(i\), and
\(\mathbf z_{ij}\) a column vector containing the row of the design
matrix of the random effects, \(\mathbf Z\), that corresponds to the
\(j\)-th measurement of subject \(i\). \(\mathbf Z\) typically contains
a subset of the variables in \(\mathbf X\), and \(\mathbf b_i\) follows
a normal distribution with mean zero and covariance matrix
\(\mathbf D\).

In both cases the parameter vector \(\boldsymbol\theta_{y\mid x}\)
contains the regression coefficients \(\boldsymbol\beta\), and
potentially additional variance parameters (e.g., for linear (mixed)
models), for which prior distributions will be specified in
Section~\ref{sec:priors}.

As mentioned, the package allows for extensions of the GLMM using a
log-normal and beta distribution, whichever is appropriate for he data
at hand. In the log-normal model, a log-normal distribution is assumed
for the outcome \(y\). This distribution is parametrized in terms of the
log scale, i.e., \(E(\log(y)) = \mathbf x_{ij}^\top\boldsymbol\beta\)
or, in case of a log-normal mixed model
\(E(\log(y)) = \mathbf x_{ij}^\top\boldsymbol\beta + \mathbf z_{ij}^\top\mathbf b_i\).

\bigskip

The beta distribution is parametrized \begin{eqnarray*}
y_{ij} & \sim & Beta(a_{ij}, b_{ij}),\\
a_{ij} &=& \mu_{ij}\; \tau,\\
b_{ij} &=& (1 - \mu_{ij})\; \tau,\\
\text{logit}(\mu_{ij}) &=& \mathbf x_{ij}^\top\boldsymbol\beta + 
\mathbf z_{ij}^\top\mathbf b_i,
\end{eqnarray*} where \(\mu_{ij}\) is the expected value of subject
\(i\) at measurement occasion \(j\),
\(\text{logit}(x) = \log\left(\frac{x}{1-x}\right)\), and \(\tau\)
follows a Gamma distribution.

\hypertarget{cumulative-logit-mixed-models}{%
\subsubsection{Cumulative logit (mixed)
models}\label{cumulative-logit-mixed-models}}

Cumulative logit mixed models are of the form \begin{eqnarray*}
y_{ij} &\sim& \text{Mult}(\pi_{ij,1}, \ldots, \pi_{ij,K}),\\[2ex]
\pi_{ij,1} &=& 1 - \sum_{k = 2}^K \pi_{ij, k},\\
\pi_{ij,k} &=& P(y_{ij} > k-1) - P(y_{ij} > k), \quad k \in 2, \ldots, K-1,\\
\pi_{ij,K} &=& P(y_{ij} \geq k-1),\\[2ex]
\text{logit}(P(y_{ij} > k)) &=& \gamma_k + \eta_{ij}, \quad k \in 1,\ldots,K,\\
\eta_{ij} &=& \mathbf x_{ij}^\top\boldsymbol\beta + \mathbf z_{ij}^\top\mathbf b_i,\\[2ex]
\gamma_1,\delta_1,\ldots,\delta_{K-1} &\overset{iid}{\sim}& N(\mu_\gamma, \sigma_\gamma^2),\\
\gamma_k &\sim& \gamma_{k-1} + \exp(\delta_{k-1}),\quad k = 2,\ldots,K,
\end{eqnarray*} where \(\pi_{ij,k} = P(y_{ij} = k)\). A cumulative logit
regression model for a univariate outcome \(y_i\) can be obtained by
dropping the index \(j\) and omitting
\(\mathbf z_{ij}^\top\mathbf b_i\). In cumulative logit (mixed) models,
the design matrix \(\mathbf X\) does not contain an intercept, since
outcome category specific intercepts \(\gamma_1,\ldots, \gamma_K\) are
specified. Here, the parameter vector \(\boldsymbol \theta_{y\mid x}\)
includes the regression coefficients \(\boldsymbol\beta\), the first
intercept \(\gamma_1\) and increments
\(\delta_1, \ldots, \delta_{K-1}\).

Note that this implementation assumes proportional odds, i.e., that the
linear predictors for the different categories of the outcome only
differ in the intercepts, but that covariates have the same effect on
the probability to be in the respective next category. This assumption
can be relaxed for some or all of the regression coefficients by
extending the linear predictor to \(\gamma_k + \eta_{ijk}\) with
\(\eta_{ijk} = \mathbf x_{ij}^\top\boldsymbol\beta_k + \mathbf z_{ij}^\top\mathbf b_i\).

\pagebreak

\hypertarget{multinomial-logit-mixed-models}{%
\subsubsection{Multinomial logit (mixed)
models}\label{multinomial-logit-mixed-models}}

Multinomial logit mixed models are implemented as \begin{eqnarray*}
y_{ij} &\sim& \text{Mult}(\pi_{ij,1}, \ldots, \pi_{ij,K}),\\[2ex]
\pi_{ij,k} &=& \phi_{i,k} / \sum_{q = 1}^K \phi_{i, q}, \quad k \in 1, \ldots, K,\\
\log(\phi_{ij, 1}) &=& 0,\\
\log(\phi_{ij, k}) &=& \mathbf x_{ij}^\top\boldsymbol\beta_k + \mathbf z_{ij}^\top\mathbf b_i,
\quad k \in 2, \ldots, K,
\end{eqnarray*} where \(\pi_{ij,k} = P(y_{ij} = k)\) is the probability
to observe category \(k\) for subject \(i\) at measurement occasion
\(j\).

\hypertarget{survival-models}{%
\subsubsection{Survival models}\label{survival-models}}

Survival data are typically characterized by the observed event or
censoring times, \(T_i\), and the event indicator, \(D_i\), which is one
if the event was observed and zero otherwise. \pkg{JointAI} provides two
types of models to analyse right censored survival data, a parametric
model which assumes a Weibull distribution for the true (but partially
unobserved) survival times \(T^*\), and a semi-parametric proportional
hazards model.

The parametric survival model is implemented as
\begin{eqnarray*} T_i^* &\sim&
\text{Weibull}(1, r_i, s),\\ D_i &\sim& \mathds{1}(T_i^* \geq C_i),\\ \log(r_j)
&=& - \mathbf x_i^\top\boldsymbol\beta,\\ s &\sim& \text{Exp}(0.01),
\end{eqnarray*} where \(\mathds{1}(T_i^* \geq C_i)\) is the indicator
function which is one if \(T_i^*\geq C_i\), and zero otherwise.

The proportional hazards model can be written as
\[h_i(t) = h_0(t)\exp(\mathbf X_i \boldsymbol\beta),\] where \(h_0(t)\)
is the baseline hazard function, which, in \pkg{JointAI}, is modelled
using a B-spline approach with \(Q\) degrees of freedom, i.e.,
\(h_0(t) = \sum_{q = 1}^\text{Q} \gamma_{Bq} B_q(t),\) where \(B_q\)
denotes the \(q\)-th basis function and \(\gamma_{Bq}\) the
corresponding regression coefficient.

The survival function of the proportional hazards model with
time-constant covariates is \[S(t\mid \boldsymbol\theta) =
\exp\left\{-\int_0^th_0(s)\exp\left(\mathbf X_i\boldsymbol\beta\right)ds\right\}
=
\exp\left\{-\exp\left(\mathbf X_i\boldsymbol\beta\right)\int_0^th_0(s)ds\right\},\]
where \(\boldsymbol\theta\) includes the regression coefficients
\(\boldsymbol\beta\) (which do not include an intercept) and the
coefficients \(\boldsymbol \gamma_{B}\) used in the specification of the
baseline hazard. Since the integral over the baseline hazard does not
have a closed-form solution, in \pkg{JointAI} it is approximated using
Gauss-Kronrod quadrature with 15 evaluation points.

\pagebreak

\hypertarget{joint-models}{%
\subsubsection{Joint models}\label{joint-models}}

Joint models for longitudinal and survival data are implemented using a
semi-parametric proportional hazards model for the time-to-event outcome
and mixed models for the longitudinal outcomes. The linear predictor of
the proportional hazards model is then
\[\exp(\mathbf X_i \boldsymbol\beta + f(s_i(t))\beta_s),\] where
\(f(s_i(t))\) is denotes a function that describes the association the
hazard has with the longitudinal variable and \(\beta_s\) is the
regression coefficient associated with it. In the simplest case, this
could be the observed or imputed value, i.e.,
\(f(s_i(t)) = \widehat s_i\), or the expected value (i.e., the value of
the linear predictor),
\(f(s_i(t)) = E(s_i\mid t, \mathbf X, \mathbf b_i, \boldsymbol\theta)\).

To take into account potential correlation between multiple time-varying
covariates, an association structure between them can be specified
explicitly by including the time-varying covariates in each other's
linear predictors in a sequential manner, or their random effects can be
modelled jointly.

\hypertarget{sec:imppart}{%
\subsection{Imputation part}\label{sec:imppart}}

A convenient way to specify the joint distribution of the incomplete
covariates
\(\mathbf X_{mis} = (\mathbf x_{mis_1}, \ldots, \mathbf x_{mis_q})\) is
to use a sequence of conditional univariate distributions
\citep{Ibrahim2002, Erler2016}

\begin{eqnarray}\label{eqn:factorization}\nonumber
p(\mathbf x_{mis_1}, \ldots, \mathbf x_{mis_q} \mid \mathbf X_{obs}, \boldsymbol\theta_{x})
& = & p(\mathbf x_{mis_1} \mid \mathbf X_{obs}, \boldsymbol\theta_{x_1})\\
&   & \prod_{\ell=2}^q p(\mathbf x_{mis_{\ell}} \mid \mathbf X_{obs}, \mathbf x_{mis_1},
\ldots, \mathbf x_{mis_{\ell-1}}, \boldsymbol\theta_{x_\ell}),
\end{eqnarray} with
\(\boldsymbol\theta_{x} = (\boldsymbol\theta_{x_1}^\top, \ldots, \boldsymbol\theta_{x_q}^\top)^\top\).

Each of the conditional distributions is a member of the exponential
family, extended with distributions for categorical variables, beta and
log-normal models, and chosen according to the type of the respective
variable. Its linear predictor is \[
g_\ell\left\{E\left(x_{i,mis_\ell} \mid \mathbf x_{i,obs},
\mathbf x_{i, mis_{<\ell}},
\boldsymbol\theta_{x_\ell}\right)
\right\} = (\mathbf x_{i, obs}^\top, x_{i, mis_1},
\ldots, x_{i, mis_{\ell-1}}) \boldsymbol\alpha_{\ell},
\quad \ell=1,\ldots,q,
\] where
\(\mathbf x_{i,mis_{<\ell}} = (x_{i,mis_1}, \ldots, x_{i,mis_{\ell-1}})^\top\)
and \(\mathbf x_{i,obs}\) is the vector of values for subject \(i\) of
those covariates that are observed for all subjects.

Factorization of the joint distribution of the covariates in such a
sequence yields a straightforward specification of the joint
distribution, even when the covariates are of mixed type. Missing values
in the covariates are sampled from their full-conditional distribution
that can be derived from the full joint distribution of outcome and
covariates. When, for instance, the analysis model is a GLM, the full
conditional distribution of an incomplete covariate
\(x_{i, mis_{\ell}}\) can be written as
\begin{eqnarray}\nonumber\label{eqn:fullcond}
p(x_{i, mis_{\ell}} \mid \mathbf y_i, \mathbf x_{i,obs},
\mathbf x_{i,mis_{-\ell}}, \boldsymbol\theta)
&\propto& p \left(y_i \mid \mathbf x_{i, obs}, \mathbf x_{i, mis},
\boldsymbol\theta_{y\mid x}
\right)
p(\mathbf x_{i, mis}\mid \mathbf x_{i, obs}, \boldsymbol\theta_{x})\,
p(\boldsymbol\theta_{y\mid x})\, p(\boldsymbol\theta_{x})\\\nonumber
&\propto& p \left(y_i \mid \mathbf x_{i, obs}, \mathbf x_{i, mis},
\boldsymbol\theta_{y\mid x}
\right)\\\nonumber
&        & p(x_{i, mis_\ell} \mid \mathbf x_{i, obs}, \mathbf x_{i, mis_{<\ell}}, \boldsymbol\theta_{x_\ell})\\\nonumber
&        & \left\{
\prod_{k=\ell+1}^q p(x_{i,mis_k}\mid \mathbf x_{i, obs},
\mathbf x_{i, mis_{<k}},
\boldsymbol\theta_{x_k})
\right\}\\
&        & p(\boldsymbol\theta_{y\mid x}) p(\boldsymbol\theta_{x_\ell})
\prod_{k=\ell+1}^p p(\boldsymbol\theta_{x_k}),
\end{eqnarray} where \(\boldsymbol\theta_{x_{\ell}}\) is the vector of
parameters describing the model for the \(\ell\)-th covariate, and
contains the vector of regression coefficients
\(\boldsymbol\alpha_\ell\) and potentially additional (e.g., variance)
parameters. The product of distributions enclosed by curly brackets
represents the distributions of those covariates that have
\(x_{mis_\ell}\) as a predictive variable in the specification of the
sequence in (\ref{eqn:factorization}).

Note that the imputed values for \(x_{i, mis_{\ell}}\) are sampled from
(\ref{eqn:fullcond}), which is the actual imputation model, and that the
conditional distributions of \(x_{i, mis_{\ell}}\) from
(\ref{eqn:factorization}) are the models that are explicitly specified
in the product that forms the joint distribution.

\hypertarget{sec:impLong}{%
\subsubsection{Imputation in multi-level settings}\label{sec:impLong}}

Factorizing the joint distribution into analysis model and imputation
part also facilitates extensions to settings with more complex outcomes,
such as repeatedly measured outcomes. In the case where the analysis
model is a mixed model with two levels, the conditional distribution of
the outcome in (\ref{eqn:fullcond}),
\(p\left(y_i \mid \mathbf x_{i, obs}, \mathbf x_{i, mis}, \boldsymbol\theta_{y\mid x} \right),\)
has to be replaced by \begin{eqnarray}\label{eqn:lmmpartfullcond}
\left\{\prod_{j=1}^{n_i} p \left(y_{ij} \mid \mathbf x_{i, obs},
\mathbf x_{i, mis}, \mathbf b_i,
\boldsymbol\theta_{y\mid x}\right)
\right\}.
\end{eqnarray} Since \(\mathbf y\) does not appear in any of the other
terms in (\ref{eqn:fullcond}), and (\ref{eqn:lmmpartfullcond}) can be
chosen to be a model that is appropriate for the outcome at hand, the
thereby specified full conditional distribution of \(x_{i, mis_\ell}\)
allows us to draw valid imputations that use all available information
on the outcome.

This is an important difference to standard FCS, where the full
conditional distributions used to impute missing values are specified
directly, usually as regression models, and require the outcome to be
explicitly included into the linear predictor of the imputation model.
In settings with complex outcomes it is not clear how this should be
done and simplifications may lead to biased results \citep{Erler2016}.
The joint model specification utilized in \pkg{JointAI} overcomes this
difficulty.

When some covariates are repeatedly measured, it is convenient to
specify models for these variables in the beginning of the sequence of
covariate models, so that models for lower level variables (e.g.,
level-1) have variables of the same or higher levels (e.g., level-1,
level-2, level-3, \ldots) in their linear predictor, but lower level
covariates do not enter the predictors of higher level covariates. Note
that, whenever there are incomplete higher level covariates it is
necessary to specify models for all lower level variables, even
completely observed ones, while models for completely observed
covariates on the highest level of the hierarchy can be omitted. This
becomes clear when we explicitly extend the factorized joint
distribution from above with completely and incompletely observed
level-1 covariates \(\mathbf s_{obs}\) and \(\mathbf s_{mis}\):
\begin{multline*} p \left(y_{ij} \mid
\mathbf s_{ij, obs}, \mathbf s_{ij, mis}, \mathbf x_{i, obs}, \mathbf x_{i,
mis}, \boldsymbol\theta_{y\mid x} \right)\\ p(\mathbf s_{ij, mis}\mid \mathbf
s_{ij, obs}, \mathbf x_{i, obs}, \mathbf x_{i, mis},
\boldsymbol\theta_{s_{mis}})\, p(\mathbf s_{ij, obs}\mid \mathbf x_{i, obs},
\mathbf x_{i, mis}, \boldsymbol\theta_{s_{obs}})\\ p(\mathbf x_{i, mis}\mid
\mathbf x_{i, obs}, \boldsymbol\theta_{x_{mis}})\, p(\mathbf x_{i, obs} \mid
\boldsymbol\theta_{x_{obs}})\, p(\boldsymbol\theta_{y\mid x})\,
p(\boldsymbol\theta_{s_{mis}}) \, p(\boldsymbol\theta_{s_{obs}})\,
p(\boldsymbol\theta_{x_{mis}}) \, p(\boldsymbol\theta_{x_{obs}})
\end{multline*} Given that the parameter vectors
\(\boldsymbol\theta_{x_{obs}}\), \(\boldsymbol\theta_{x_{mis}}\),
\(\boldsymbol\theta_{s_{obs}}\) and \(\boldsymbol\theta_{s_{mis}}\) are
a priori independent, and
\(p(\mathbf x_{i, obs} \mid \boldsymbol\theta_{x_{obs}})\) is
independent of both \(\mathbf x_{mis}\) and \(\mathbf s_{mis}\), it can
be omitted.

Since
\(p(\mathbf s_{ij, obs}\mid \mathbf x_{i, obs}, \mathbf x_{i, mis}, \boldsymbol\theta_{s_{obs}})\),
however, has \(\mathbf x_{i, mis}\) in its linear predictor and will,
hence, be part of the full conditional distribution of
\(\mathbf x_{i, mis}\), it cannot be omitted from the model, unless it
is reasonable to assume that \(\mathbf x_{i, mis}\) and
\(\mathbf s_{ij, obs}\) are independent.

\hypertarget{non-linear-associations-and-interactions}{%
\subsubsection{Non-linear associations and
interactions}\label{non-linear-associations-and-interactions}}

Other settings in which the fully Bayesian approach employed in
\pkg{JointAI} has an advantage over standard FCS are settings with
interaction terms that involve incomplete covariates or when the
association of the outcome with an incomplete covariate is non-linear.
In standard FCS such settings lead to incompatible imputation models
\citep{White2011, Bartlett2015}. This becomes clear when considering the
following simple example where the analysis model of interest is the
linear regression
\(y_i = \beta_0 + \beta_1 x_i + \beta_2 x_i^2 + \varepsilon_i\) and
\(x_i\) is imputed using
\(x_i = \alpha_0 + \alpha_1 y_i + \tilde\varepsilon_i\). While the
analysis model assumes a quadratic relationship, the imputation model
assumes a linear association between \(\mathbf x\) and \(\mathbf y\) and
there cannot be a joint distribution that has the imputation and
analysis model as its full conditional distributions.

Because, in \pkg{JointAI}, the analysis model is a factor in the full
conditional distribution that is used to impute \(x_i\), the non-linear
association is taken into account. Furthermore, since it is the joint
distribution that is specified, and the full conditional then derived
from it, the joint distribution is ensured to exist.

\hypertarget{sec:priors}{%
\subsection{Prior distributions}\label{sec:priors}}

Prior distributions have to be specified for all (hyper)parameters. A
common prior choice for the regression coefficients is the normal
distribution with mean zero and large variance. In \pkg{JointAI},
variance parameters are specified as, by default vague, inverse-gamma
distributions.

The covariance matrix of the random effects in a mixed model,
\(\mathbf D\), is assumed to follow an inverse Wishart distribution
where the degrees of freedom are, by default, chosen to be the dimension
of the random effects plus one, and the scale matrix is diagonal. Since
the magnitude of the diagonal elements relates to the variance of the
random effects, the choice of suitable values depends on the scale of
the variable the random effect is associated with. Therefore,
\pkg{JointAI} uses independent gamma hyper-priors for each of the
diagonal elements. More details about the default hyper-parameters and
how to change them are given in Section~\ref{sec:hyperpars} and
Appendix~\ref{sec:AppHyperpars}.

\pagebreak

\hypertarget{sec:package}{%
\section{Package structure}\label{sec:package}}

The package \pkg{JointAI} has several main functions, \code{lm_imp()},
\code{glm_imp()}, \code{clm_imp()}, \ldots, abbreviated \code{*_imp()},
that perform regression of continuous and categorical, univariate or
multi-level data as well as right-censored survival data. The model
specification is similar to the specification of standard regression
models in \proglang{R} and described in detail in
Section~\ref{sec:ModelSpecification}.

Based on the specified model formula and other function arguments,
\pkg{JointAI} does some pre-processing of the data. It checks which
variables are incomplete and identifies their measurement level and
level in the hierarchical structure in order to specify appropriate
(imputation) models. Interactions and functional forms of variables are
detected in the model formula, and the design matrices for the various
parts of the model are created.

MCMC sampling is performed by the program \proglang{JAGS} \citep{JAGS}.
The \proglang{JAGS} model, data list, containing all necessary parts of
the data, and user-specified settings for the MCMC sampling (see
Section~\ref{sec:MCMCSettings}) are passed to \proglang{JAGS} via the
\proglang{R} package \pkg{rjags} \citep{rjags}.

The main functions \code{*_imp()} all return an object of class
\code{JointAI}. Summary and plotting methods for \code{JointAI} objects,
as well as functions to evaluate convergence and precision of the MCMC
samples, to predict from \code{JointAI} objects and to export imputed
values are discussed in Section~\ref{sec:Results}.

Currently, the package works under the assumption of a missing at random
(MAR) missingness process \citep{Rubin1976, Rubin1987}. When this
assumption holds, observations with missing outcome may be excluded from
the analysis in the Bayesian framework. Hence, missing values in the
outcome do not require special treatment in this setting, and,
therefore, our focus here is on missing values in covariates.
Nevertheless, \pkg{JointAI} can handle missing values in the outcome;
they are automatically imputed using the specified analysis model.

\hypertarget{example-data}{%
\section{Example data}\label{example-data}}

To illustrate the functionality of \pkg{JointAI}, we use three datasets
that are part of this package. The \code{NHANES} data contain
measurements from a cross-sectional cohort study, whereas the
\code{simLong} data is a simulated dataset based on a longitudinal
cohort study in toddlers. The third dataset (\code{PBC}) is the well
known data on primary biliary cirrhosis from the Mayo clinic.

\hypertarget{the-nhanes-data}{%
\subsection{The NHANES data}\label{the-nhanes-data}}

The \code{NHANES} data is a subset of observations from the 2011 -- 2012
wave of the National Health and Nutrition Examination Survey
\citep{NHANES2011} and contains information on 186 men and women between
20 and 80 years of age. The variables contained in this dataset are

\begin{itemize}
\tightlist
\item
  \code{SBP}: systolic blood pressure in mmHg; complete
\item
  \code{gender}: \code{male} vs \code{female}; complete
\item
  \code{age}: in years; complete
\item
  \code{race}: 5 unordered categories; complete
\item
  \code{WC}: waist circumference in cm; 1.1\% missing
\item
  \code{alc}: weekly alcohol consumption; binary; 18.3\% missing
\item
  \code{educ}: educational level; binary; complete
\item
  \code{creat}: creatinine concentration in mg/dL; 4.5\% missing
\item
  \code{albu}: albumin concentration in g/dL; 4.3\% missing
\item
  \code{uricacid}: uric acid concentration in mg/dL; 4.3\% missing
\item
  \code{bili}: bilirubin concentration in mg/dL; 4.3\% missing
\item
  \code{occup}: occupational status; 3 unordered categories; 15.1\%
  missing
\item
  \code{smoke}: smoking status; 3 ordered categories; 3.8\% missing
\end{itemize}

Figure~\ref{fig:distrplotNHANES} shows histograms and bar plots of all
continuous and categorical variables, respectively, together with the
proportion of missing values for incomplete variables. Such a plot can
be obtained with the function \code{plot_all()}. Arguments \code{fill}
and \code{border} allow the user to change colours, the number of rows
and columns can be adapted using \code{nrow} and/or \code{ncol}, and
additional arguments can be passed to \code{hist()} and \code{barplot()}
via \code{"..."}.

\begin{CodeChunk}
\begin{figure}

{\centering \includegraphics[width=1\linewidth]{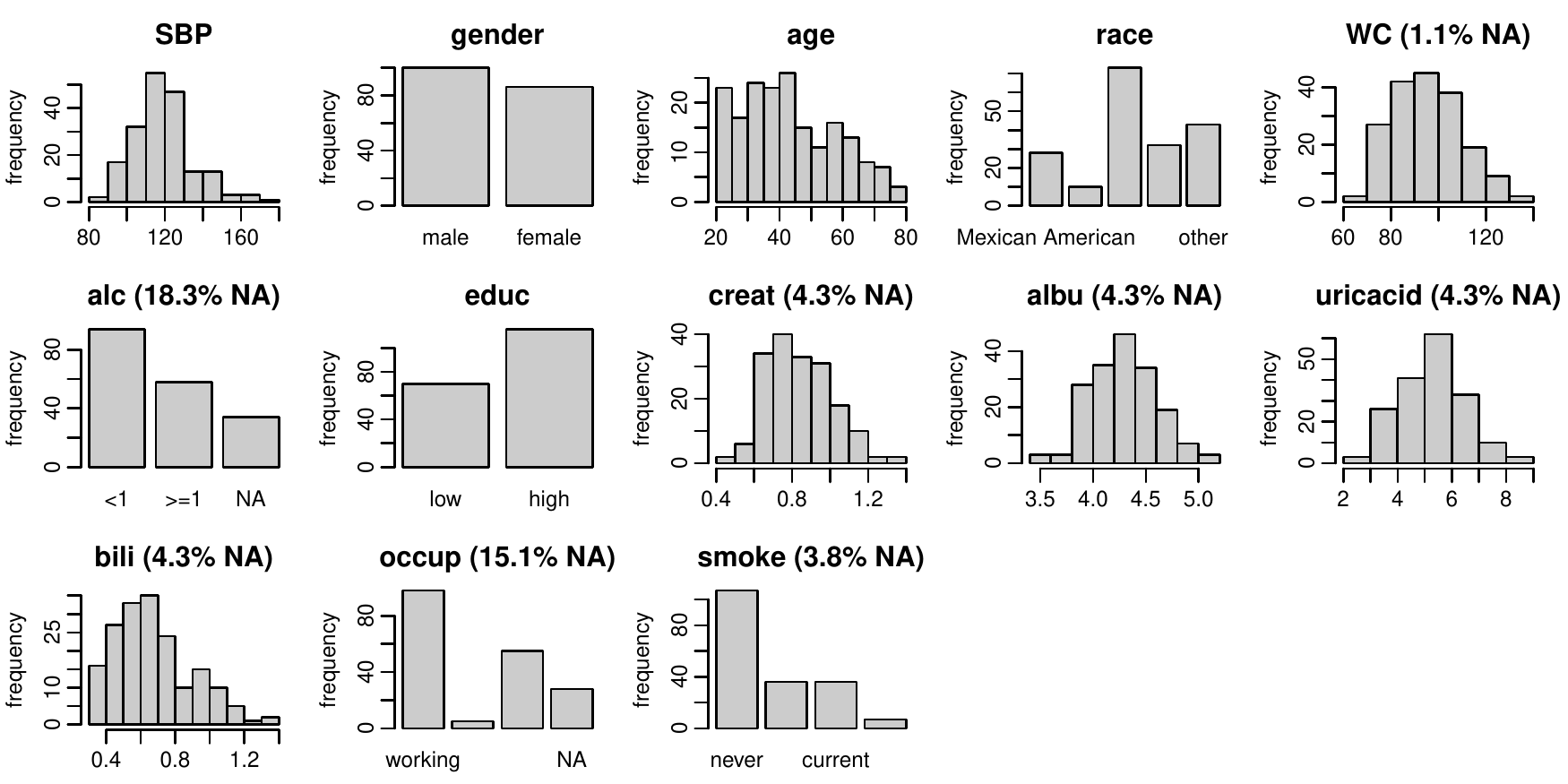} 

}

\caption[Distribution of the variables in the \code{NHANES} data (with 
percentage of missing values given for incomplete variables)]{Distribution of the variables in the \code{NHANES} data (with 
percentage of missing values given for incomplete variables).}\label{fig:distrplotNHANES}
\end{figure}
\end{CodeChunk}

The pattern of missing values in the \code{NHANES} data is shown in
Figure~\ref{fig:mdpatternNHANES}. This plot can be obtained using the
function \code{md_pattern()}. Again, arguments \code{color} and
\code{border} allow the user to change colours and arguments such as
\code{legend.position}, \code{print_xaxis} and \code{print_yaxis} permit
further customization.

\begin{CodeChunk}
\begin{figure}[!ht]

{\centering \includegraphics[width=0.8\linewidth]{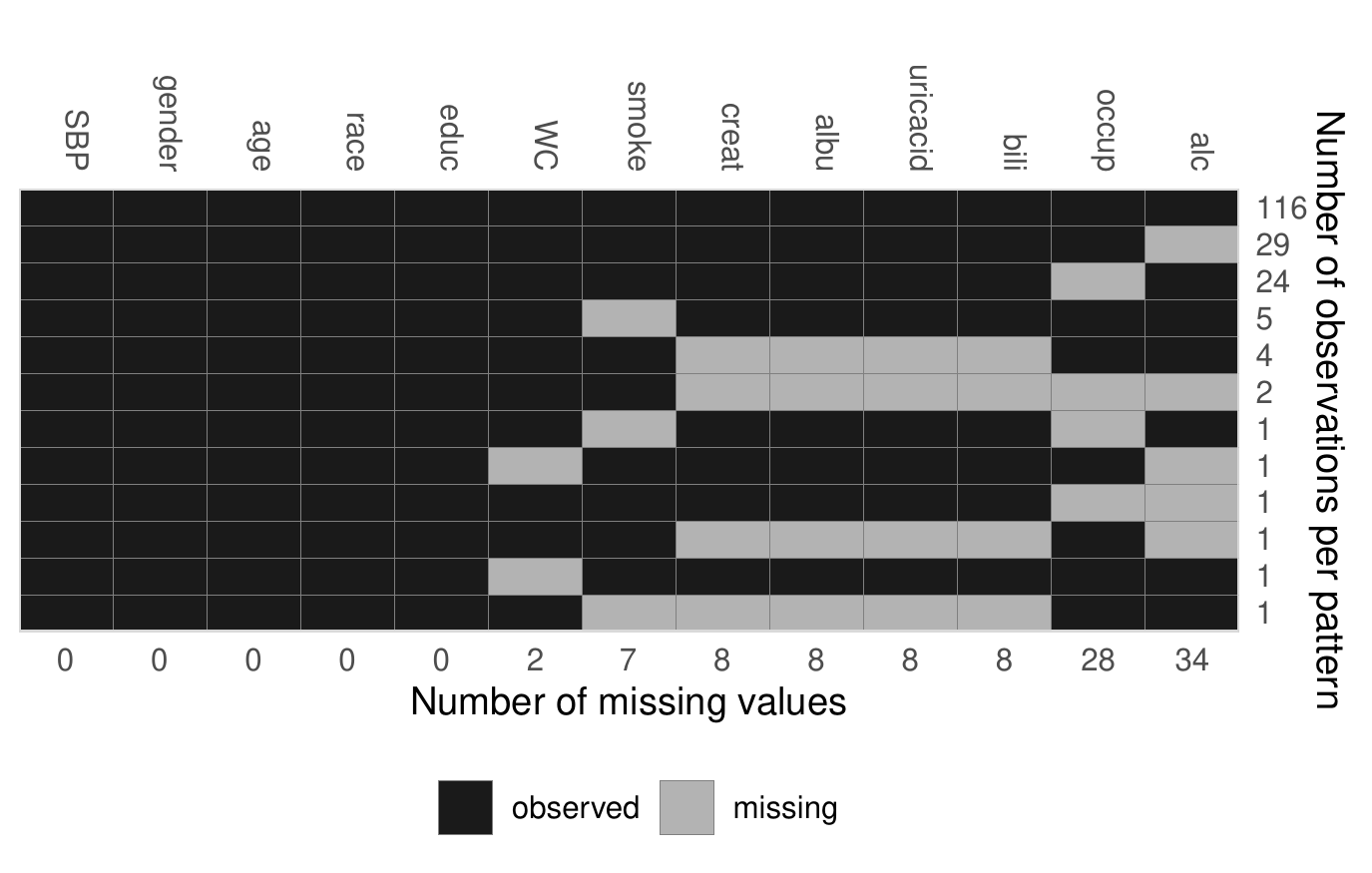} 

}

\caption[Missing data pattern of the \code{NHANES} data]{Missing data pattern of the \code{NHANES} data.}\label{fig:mdpatternNHANES}
\end{figure}
\end{CodeChunk}

Each row represents a pattern of missing values, where observed
(missing) values are depicted with dark (light) colour. The frequency
with which each of the patterns is observed is given on the right
margin, the number of missing values in each variable is given
underneath the plot. Rows and columns are ordered by number of cases per
pattern (decreasing) and number of missing values (increasing). The
first row, for instance, shows that there are 116 complete cases, the
second row that there are 29 cases for which only \code{alc} is missing.
Furthermore, it is apparent that \code{creat}, \code{albu},
\code{uricacid} and \code{bili} are always missing together. Since these
variables are all measured in serum, this is not surprising.

\code{md_pattern()} also returns the missing data pattern in matrix
representation (\code{pattern = TRUE}), where missing and observed
values are represented with a \code{0} and \code{1}, respectively.

\FloatBarrier

\hypertarget{the-simlong-data}{%
\subsection{The simLong data}\label{the-simlong-data}}

The \code{simLong} data is a simulated dataset mimicking a longitudinal
cohort study of 200 mother-child pairs. It contains the following
baseline (i.e., not time-varying) covariates\vspace*{-1ex}

\begin{itemize}
\tightlist
\item
  \code{GESTBIR}: gestational age at birth in weeks; complete
\item
  \code{ETHN}: ethnicity; binary; 2.8\% missing
\item
  \code{AGE_M}: age of the mother at intake; complete
\item
  \code{HEIGHT_M}: height of the mother in cm; 2.0\% missing
\item
  \code{PARITY}: number of times the mother has given birth; binary;
  2.4\% missing
\item
  \code{SMOKE}: smoking status of the mother during pregnancy; 3 ordered
  categories; 12.2\% missing
\item
  \code{EDUC}: educational level of the mother; 3 ordered categories;
  7.8\% missing
\item
  \code{MARITAL}: marital status; 3 unordered categories; 7.0\% missing
\item
  \code{ID}: subject identifier
\end{itemize}

and seven longitudinal variables:\vspace*{-2ex}

\begin{itemize}
\tightlist
\item
  \code{time}: measurement occasion/visit (by design, children should
  have been measured at/around 1, 2, 3, 4, 7, 11, 15, 20, 26, 32, 40 and
  50 months of age)
\item
  \code{age}: child's age at measurement time in months
\item
  \code{hgt}: child's height in cm; 20\% missing
\item
  \code{wgt}: child's weight in gram; 8.8\% missing
\item
  \code{bmi}: child's BMI (body mass index) in kg/m\textsuperscript{2};
  21.6\% missing
\item
  \code{hc}: child's head circumference in cm; 23.6\% missing
\item
  \code{sleep}: child's sleeping behaviour; 3 ordered categories; 24.7\%
  missing
\end{itemize}

Figure~\ref{fig:trajectories} shows the longitudinal profiles of
\code{hgt}, \code{wgt}, \code{bmi} and \code{hc} over age. All four
variables clearly have a non-linear pattern over time.

\begin{CodeChunk}
\begin{figure}[!ht]

{\centering \includegraphics[width=1\linewidth]{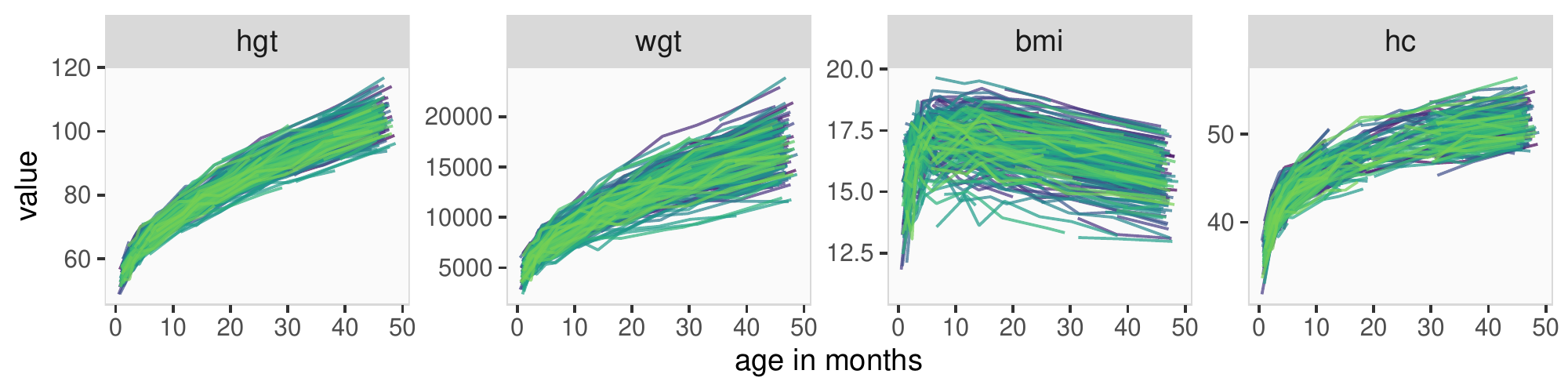} 

}

\caption[Trajectories of the continuous time-varying variables in the \code{simLong} data]{Trajectories of the continuous time-varying variables in the \code{simLong} data.}\label{fig:trajectories}
\end{figure}
\end{CodeChunk}

Histograms and bar plots of all variables in the \code{simLong} data are
displayed in Figure~\ref{fig:distrplotsimLong}. Here, the argument
\code{idvar} of the function \code{plot_all()} is used to display
baseline (level-2) covariates on the subject level instead of the
observation level:

\begin{CodeChunk}

\begin{CodeInput}
R> plot_all(simLong, use_level = TRUE, idvar = "ID", ncol = 5)
\end{CodeInput}
\end{CodeChunk}

\begin{CodeChunk}
\begin{figure}[!ht]

{\centering \includegraphics[width=1\linewidth]{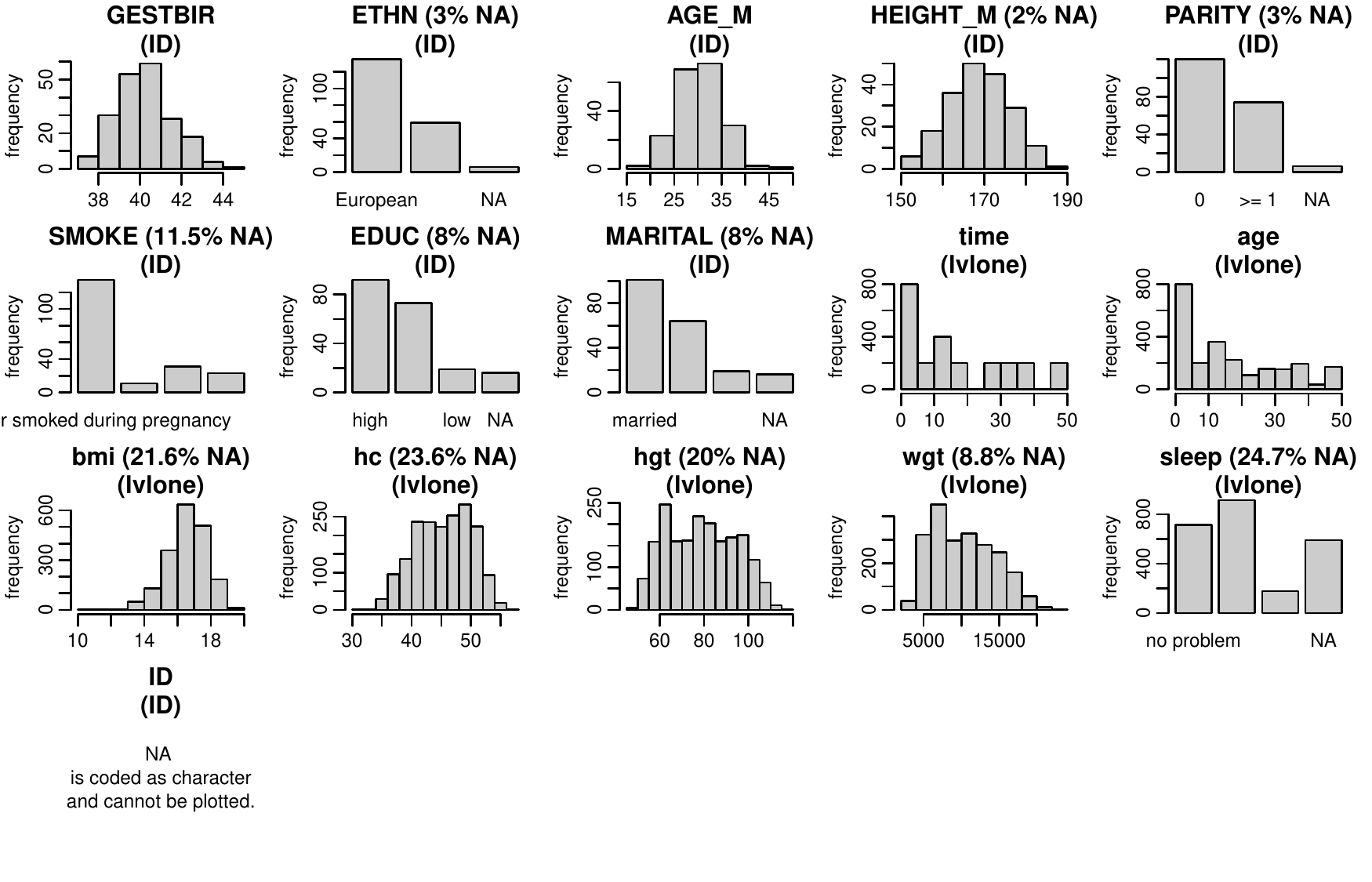} 

}

\caption[Distribution of the variables in the \code{simLong} data]{Distribution of the variables in the \code{simLong} data.}\label{fig:distrplotsimLong}
\end{figure}
\end{CodeChunk}

\begin{figure}[!ht]
\centering
\includegraphics[width = \linewidth]{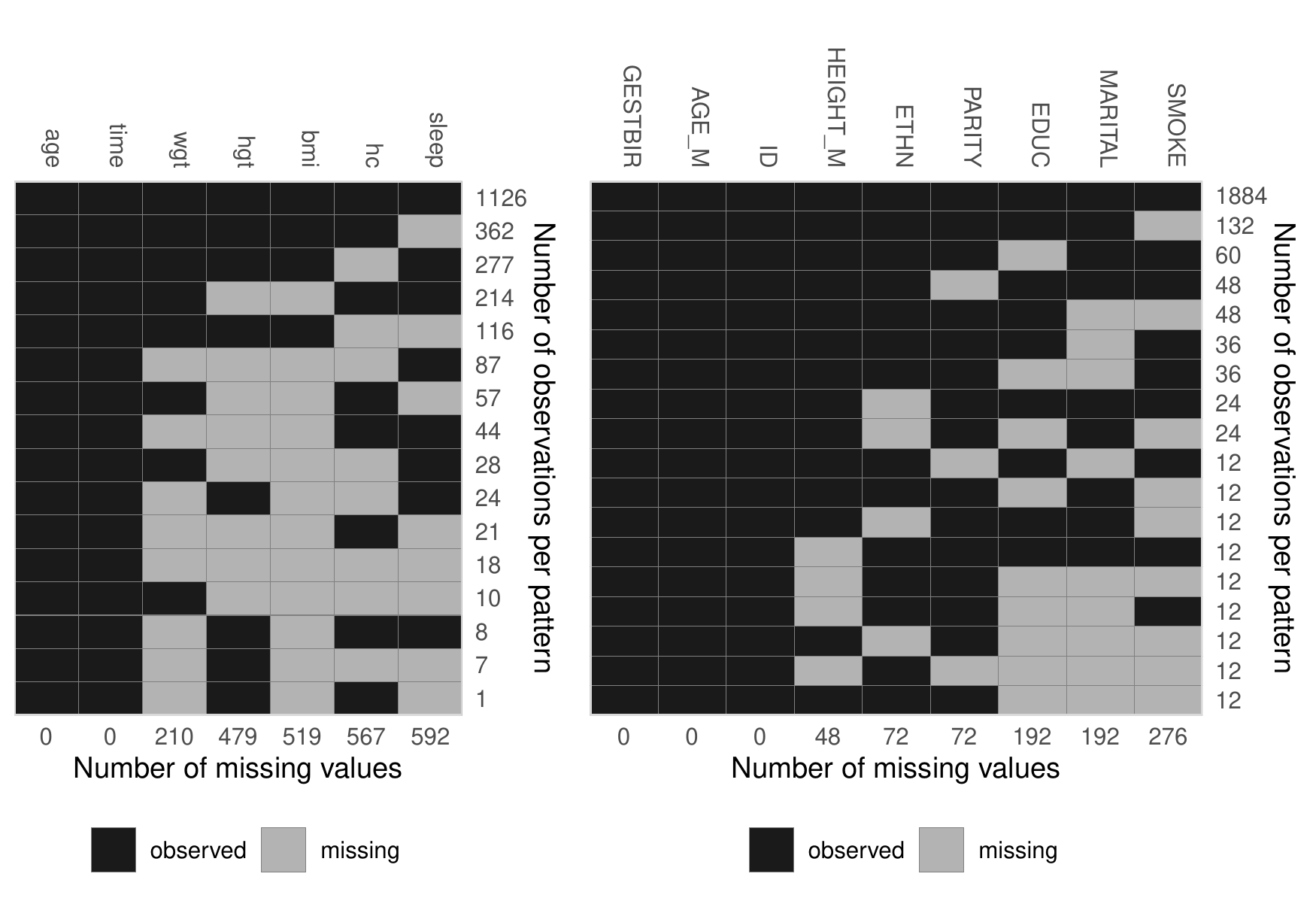}
\caption{Missing data pattern of the \code{simLong} data (left: level-1 variables, right: level-2 variables).}
\label{fig:mdpatternsimLong}
\end{figure}

The missing data pattern of the \code{simLong} data is shown in
Figure~\ref{fig:mdpatternsimLong}. For readability, the pattern is given
separately for the level-1 (left) and level-2 (right) variables. It is
non-monotone and does not have any distinctive features.

\pagebreak

\hypertarget{the-pbc-data}{%
\subsection{The PBC data}\label{the-pbc-data}}

For demonstration of the use of \pkg{JointAI} for the analysis of
survival data we use the dataset \code{PBC} which is a re-coded version
of the PBC data in the \pkg{survival} package. It contains baseline and
follow-up data of 312 patients with primary biliary cirrhosis and
includes the following variables:

Baseline covariates:\vspace*{-2ex}

\begin{itemize}
\tightlist
\item
  \code{id}: patient identifier; complete
\item
  \code{futime}: time until death, transplantation or censoring in days;
  complete
\item
  \code{status}: event indicator (\code{censored}, \code{transplant} or
  \code{dead}); complete
\item
  \code{trt}: treatment (D-penicillamine or placebo); complete
\item
  \code{age}: patient's age in years; complete
\item
  \code{sex}: patient's sex; complete
\item
  \code{copper}: urine copper (\(\mu\)g/day); 0.6\% missing
\item
  \code{trig}: triglyceride (mg/dl); 9.6\% missing
\end{itemize}

Time-varying covariates:\vspace*{-2ex}

\begin{itemize}
\tightlist
\item
  \code{day}: number of days between enrolment and this visit date (time
  variable for the laboratory measurements); complete
\item
  \code{albumin}: serum albumin (mg/dl); complete
\item
  \code{alk.phos}: alkaline phosphatase (U/litre); 3.1\% missing
\item
  \code{ascites}: presence of ascites; 3.1\% missing
\item
  \code{ast}: aspartate aminotransferase (U/ml); complete
\item
  \code{bili}: serum bilirubin (mg/dl); complete
\item
  \code{chol}: serum cholesterol (mg/dl); 42.2\% missing
\item
  \code{edema}: ``no'': no oedema, ``(un)treated'': untreated or
  successfully treated 1 oedema, ``edema'': oedema despite diuretic
  therapy; complete
\item
  \code{hepato}: presence of hepatomegaly (enlarged liver); 3.1\%
  missing
\item
  \code{platelet}: platelet count; 3.8\% missing
\item
  \code{protime}: standardised blood clotting time; complete
\item
  \code{spiders}: blood vessel malformations in the skin; 3.0\% missing
\item
  \code{stage}: histologic stage of disease (4 levels); complete
\end{itemize}

\begin{CodeChunk}
\begin{figure}

{\centering \includegraphics[width=1\linewidth]{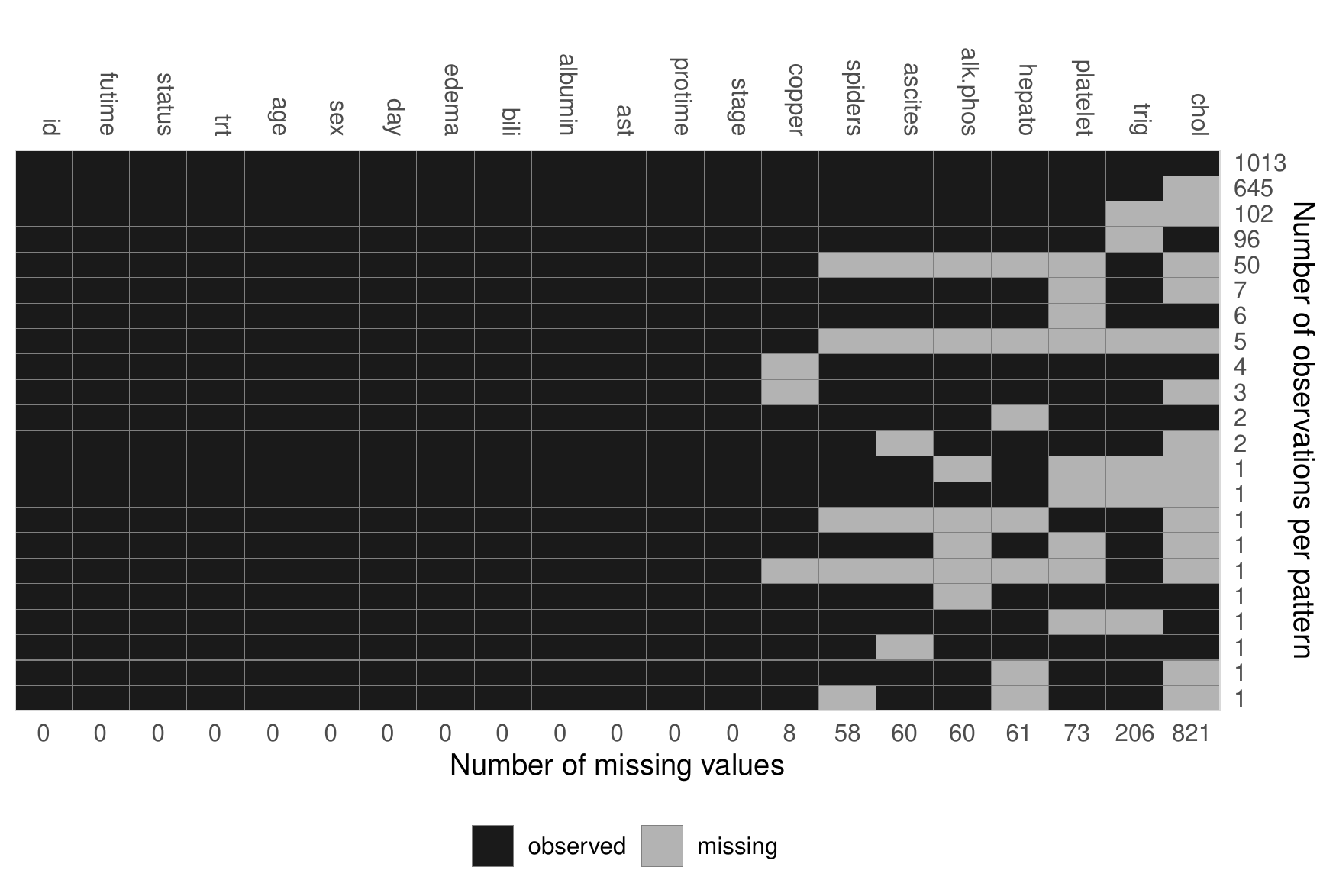} 

}

\caption[Missing data pattern of the \code{PBC} data]{Missing data pattern of the \code{PBC} data.}\label{fig:mdpatternPBC}
\end{figure}
\end{CodeChunk}

\begin{CodeChunk}
\begin{figure}[!ht]

{\centering \includegraphics[width=1\linewidth]{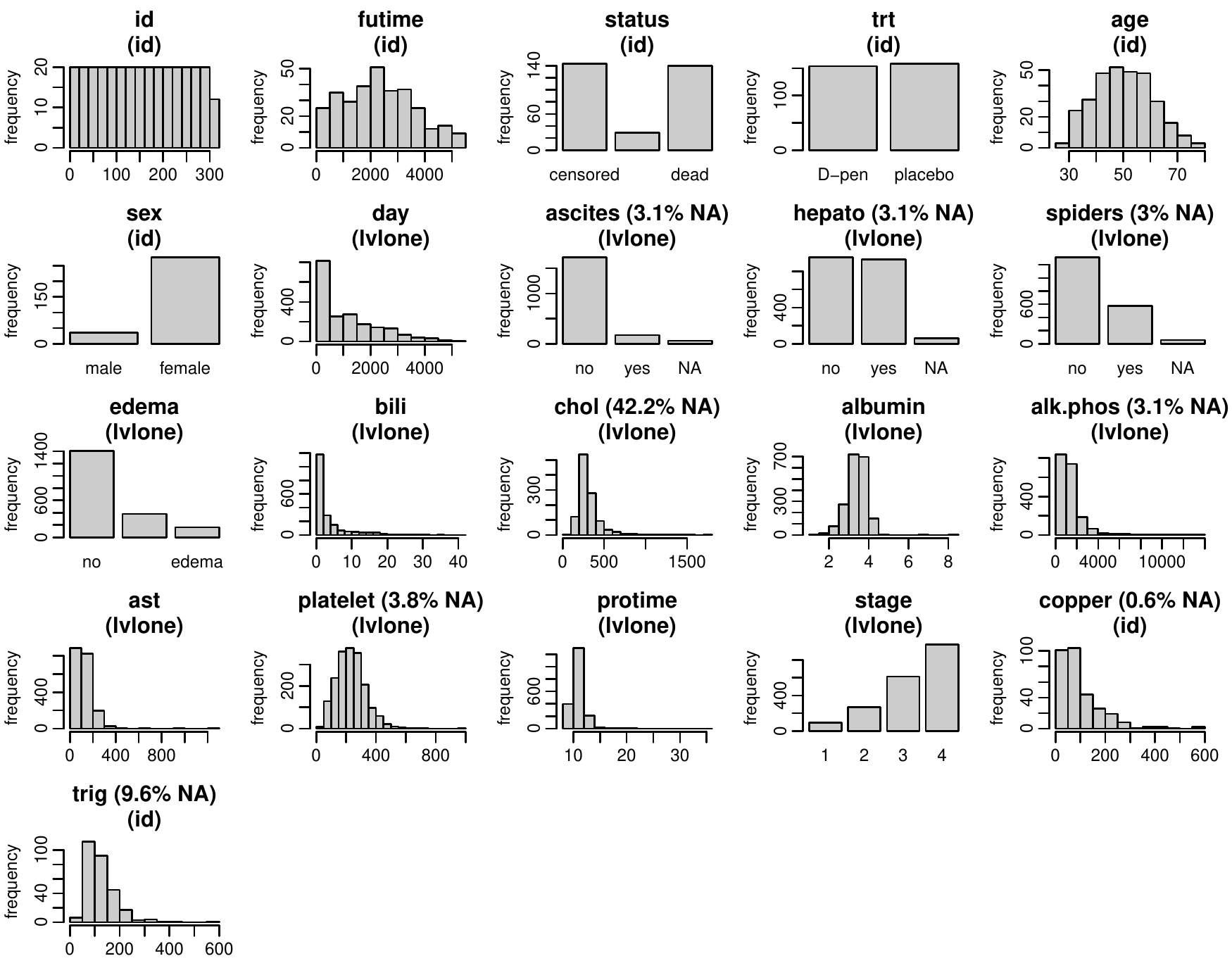} 

}

\caption[Distribution of the variables in the \code{PBC} data (with 
percentage of missing values given for incomplete variables)]{Distribution of the variables in the \code{PBC} data (with 
percentage of missing values given for incomplete variables).}\label{fig:distrplotPBC}
\end{figure}
\end{CodeChunk}

The missing data pattern and distribution of the observed values of the
\code{PBC} data is shown in Figures~\ref{fig:mdpatternPBC} and
\ref{fig:distrplotPBC}.

\hypertarget{missing-data-visualization-and-exploration}{%
\subsection{Missing data visualization and
exploration}\label{missing-data-visualization-and-exploration}}

There are several \proglang{R} packages that provide functionality for a
more in-depth exploration of incomplete data, see for example the ones
listed in the CRAN task view on missing data
(\url{https://CRAN.R-project.org/view=MissingData}). Particularly useful
may be the packages \pkg{naniar} \citep{naniar} and \pkg{VIM}
\citep{VIM}.

\FloatBarrier

\hypertarget{sec:ModelSpecification}{%
\section{Model specification}\label{sec:ModelSpecification}}

The main analysis functions in \pkg{JointAI} are \code{lm_imp()},
\code{glm_imp()}, \code{lognorm_imp()}, \code{betareg_imp()},
\code{clm_imp()}, \code{mlogit_imp()}, \code{lme_imp()},
\code{glme_imp()}, \code{lognormmm_imp()}, \code{betamm_imp()},
\code{clmm_imp()}, \code{mlogitmm_imp()}, \code{survreg_imp()},
\code{coxph_imp()} and \code{JM_imp()}.

The main arguments of these functions, i.e., \code{formula},
\code{data}, \code{family}, \code{fixed}, and \code{random}, are used
analogously to the specification in the standard complete data functions
\code{lm()} and \code{glm()} from package \pkg{stats}, \code{lme()},
from package \pkg{nlme} \citep{nlme} and \code{survreg()} and
\code{coxph()} from package \pkg{survival}, for example:

\begin{CodeChunk}

\begin{CodeInput}
R> lm_imp(formula, data, 
+    n.chains = 3, n.adapt = 100, n.iter = 0, thin = 1, ...)
 
R> glm_imp(formula, family, data,
+    n.chains = 3, n.adapt = 100, n.iter = 0, thin = 1, ...)

R> lme_imp(fixed, data, random,
+    n.chains = 3, n.adapt = 100, n.iter = 0, thin = 1, ...)
 
R> glme_imp(fixed, data, random, family,
+    n.chains = 3, n.adapt = 100, n.iter = 0, thin = 1, ...)

R> survreg_imp(formula, data,
+    n.chains = 3, n.adapt = 100, n.iter = 0, thin = 1, ...)
\end{CodeInput}
\end{CodeChunk}

The specification for \code{lognorm_imp()}, \code{betareg_imp()}, and
\code{mlogit_imp()} is the same as for \code{lm_imp()}.

The functions \code{lme_imp()} and \code{glme_imp()} have aliases
\code{lmer_imp()} and \code{glmer_imp()}, and all mixed model functions
accept specification of a combined fixed and random effects formula
(like in the package \pkg{lme4} \citep{lme4}) as well as the
specification using \code{fixed} and \code{random}.

The arguments \code{formula} and \code{fixed} take a standard two-sided
\code{formula} object, where an intercept is added automatically (except
in ordinal and proportional hazards models). For the specification of
random effects formulas, see Section~\ref{sec:multilevel}.

\code{clm_imp()} and \code{clmm_imp()} have additional optional
arguments \code{nonprop} and \code{rev}. \code{nonprop} expects a
one-sided formula containing those terms of \code{formula} or
\code{fixed} that should have non-proportional effects, and \code{rev}
can be set to \code{TRUE} to indicate that the odds should be reversed,
i.e., to model \(\frac{P(y \leq k)}{P(y > k)}\) instead of
\(\frac{P(y > k)}{P(y \leq k)}\).

Survival models expect the left hand side of \code{formula} to be a
survival object (created with the function \code{Surv()} from package
\pkg{survival}, see Section~\ref{sec:survmod}).

The argument \code{family} enables the choice of a distribution and link
function from a range of options when using \code{glm_imp()} or
\code{glme_imp()}. The implemented options are given in
Table~\ref{tab:distrlink}.

\begin{table}[!ht]
\centering
\begin{tabular}{ll}\toprule
distribution & link\\\midrule
{gaussian} & identity, log, inverse\\
{binomial} & logit, probit, log, cloglog\\
{Gamma} & inverse, identity, log\\
{poisson} & log, identity\\\bottomrule
\end{tabular}
\caption{Possible choices for the model \code{family} and \code{link} functions
in \code{glm\_imp()} and \code{glme\_imp()}.}
\label{tab:distrlink}
\end{table}

For the description of the remaining arguments see below and
Section~\ref{sec:MCMCSettings}.

\hypertarget{specification-of-the-model-formula}{%
\subsection{Specification of the model
formula}\label{specification-of-the-model-formula}}

\hypertarget{interactions}{%
\subsubsection{Interactions}\label{interactions}}

In \pkg{JointAI}, interactions between any type of variables (observed,
incomplete, variables from different hierarchical levels) can be
handled. When an incomplete variable is involved, the interaction term
is re-calculated within each iteration of the MCMC sampling, using the
imputed values from the current iteration. Interaction terms involving
incomplete variables should, hence, not be pre-calculated as an
additional variable since this would lead to inconsistent imputed values
of main effect and interaction term.

Interactions between multiple variables can be specified using
parentheses; for higher lever interactions the \code{\textasciicircum}
operator can be used:

\begin{CodeChunk}

\begin{CodeInput}
R> mod1a <- glm_imp(educ ~ gender * (age + smoke + creat),
+   data = NHANES, family = binomial())
\end{CodeInput}
\end{CodeChunk}

\begin{CodeChunk}

\begin{CodeInput}
R> mod1b <- glm_imp(educ ~ gender + (age + smoke + creat)^3,
+   data = NHANES, family = binomial())
\end{CodeInput}
\end{CodeChunk}

\hypertarget{non-linear-functional-forms}{%
\subsubsection{Non-linear functional
forms}\label{non-linear-functional-forms}}

In practice, associations between outcome and covariates do not always
meet the standard assumption of linearity. Often, assuming a
logarithmic, quadratic or other non-linear effect is more appropriate.

For completely observed covariates, \pkg{JointAI} can handle any common
type of function implemented in \proglang{R}, including splines, e.g.,
using \code{ns()} or \code{bs()} from the package \pkg{splines}. Since
functions involving variables that have missing values need to be
re-calculated in each iteration of the MCMC sampling, currently, only
functions that are available in \proglang{JAGS} can be used for
incomplete variables. Those functions include:

\begin{itemize}
\item
  \code{log()}, \code{exp()}
\item
  \code{sqrt()}, polynomials (using \code{I()})
\item
  \code{abs()}
\item
  \code{sin()}, \code{cos()}
\item
  algebraic operations (wrapped in \code{I()}) involving one or multiple
  (in)complete variables, as long as the formula can be interpreted by
  \proglang{JAGS}.
\end{itemize}

The list of functions implemented in \proglang{JAGS} can be found in the
\proglang{JAGS} user manual \citep{JAGSmanual} available at
\url{https://sourceforge.net/projects/mcmc-jags/files/Manuals/}.

Some examples (that do not necessarily have a meaningful interpretation
or good model fit) are:

\begin{CodeChunk}

\begin{CodeInput}
R> mod2a <- lm_imp(SBP ~ age + gender + abs(bili - creat), data = NHANES)
\end{CodeInput}
\end{CodeChunk}

\begin{CodeChunk}

\begin{CodeInput}
R> library("splines")
R> mod2b <- lm_imp(SBP ~ ns(age, df = 2) + gender + I(bili^2) + I(bili^3),
+   data = NHANES)
\end{CodeInput}
\end{CodeChunk}

\begin{CodeChunk}

\begin{CodeInput}
R> mod2c <- lm_imp(SBP ~ age + gender + I(creat/albu^2), data = NHANES,
+   trunc = list(albu = c(1e-5, NA)))
\end{CodeInput}
\end{CodeChunk}

\begin{CodeChunk}

\begin{CodeInput}
R> mod2d <- lm_imp(SBP ~ bili + sin(creat) + cos(albu), data = NHANES)
\end{CodeInput}
\end{CodeChunk}

It is also possible to nest a function in another function.

\begin{CodeChunk}

\begin{CodeInput}
R> mod2e <- lm_imp(SBP ~ age + gender + sqrt(exp(creat)/2), data = NHANES)
\end{CodeInput}
\end{CodeChunk}

\hypertarget{sec:restsupport}{%
\subsubsection{Functions with restricted
support}\label{sec:restsupport}}

When a function of an incomplete variable has restricted support, e.g.,
\(\log(x)\) is only defined for \(x > 0\), or, in \code{mod2c} from
above \code{I(creat/albu\textasciicircum 2)} can not be calculated for
\code{albu = 0}, the model specified for that incomplete variable needs
to comply with these restrictions. This can either be achieved by
truncating the distribution, using the argument \code{trunc}, or by
selecting a distribution that meets the restrictions.

Note that truncation should be used with care. Its intended use here is
to prevent issues when theoretically a variable could take a value that
would result in an invalid mathematical expression. Truncation should
not be used to make symmetric distributions, like the normal
distribution, fit skewed data.
\citep{vonHippel2012, Rodwell2014, Geraci2018}

\hypertarget{example}{%
\paragraph{Example:}\label{example}}

When using a \(\log\) transformation for the covariate \code{uricacid},
we can use the default imputation method \code{"norm"} (a normal
distribution) and truncate it by specifying
\code{trunc = list(uricacid = c(<lower>, <upper>))}, where
\code{<lower>} and \code{<upper>} are the smallest and largest values
allowed:

\begin{CodeChunk}

\begin{CodeInput}
R> mod3a <- lm_imp(SBP ~ age + gender + log(uricacid) + exp(creat),
+   trunc = list(uricacid = c(1e-5, NA)), data = NHANES)
\end{CodeInput}
\end{CodeChunk}

One-sided truncation is possible by setting the limit that is not needed
to \code{NA}.

Alternatively, we may choose a model for the incomplete variable (using
the argument \code{models}; for more details see
Section~\ref{sec:methJSS}) that only imputes positive values such as a
log-normal distribution or a gamma distribution:

\begin{CodeChunk}

\begin{CodeInput}
R> mod3b <- lm_imp(SBP ~ age + gender + log(uricacid) + exp(creat),
+   models = c(uricacid = "lognorm"), data = NHANES)
\end{CodeInput}
\end{CodeChunk}

\begin{CodeChunk}

\begin{CodeInput}
R> mod3c <- lm_imp(SBP ~ age + gender + log(uricacid) + exp(creat),
+   models = c(uricacid = "glm_gamma_inverse"), data = NHANES)
\end{CodeInput}
\end{CodeChunk}

\hypertarget{functions-that-are-not-available-in-r}{%
\subsubsection{Functions that are not available in
R}\label{functions-that-are-not-available-in-r}}

It is possible to use functions that have different names in
\proglang{R} and \proglang{JAGS}, or that do exist in \proglang{JAGS},
but not in \proglang{R}, by defining a new function in \proglang{R} that
has the name of the function in \proglang{JAGS}.

\hypertarget{example-1}{%
\paragraph{Example:}\label{example-1}}

In \proglang{JAGS} the inverse logit transformation is defined in the
function \code{ilogit}. In base \proglang{R}, there is no function
\code{ilogit}, but the inverse logit is available as the distribution
function of the logistic distribution \code{plogis()}. Thus, we can
define the function \code{ilogit()} as

\begin{CodeChunk}

\begin{CodeInput}
R> ilogit <- plogis
\end{CodeInput}
\end{CodeChunk}

and use it in the model formula

\begin{CodeChunk}

\begin{CodeInput}
R> mod4a <- lm_imp(SBP ~ age + gender + ilogit(creat), data = NHANES)
\end{CodeInput}
\end{CodeChunk}

\hypertarget{a-note-on-what-happens-inside-jointai}{%
\subsubsection{A note on what happens inside
JointAI}\label{a-note-on-what-happens-inside-jointai}}

When a function of a complete or incomplete variable is used in the
model formula, the main effect of that variable is automatically added
as an auxiliary variable (more on auxiliary variables in
Section~\ref{sec:auxvars}), and only the main effects are used as
predictors in the imputation models.

In \code{mod2b}, for example, the spline of \code{age} is used as
predictor for \code{SBP}, but in the imputation model for \code{bili},
\code{age} enters with a linear effect. This can be checked using the
function \code{list_models()}, which prints a list of all sub-models
used in an \pkg{JointAI} model. Here, we are only interested in the
predictor variables, and, hence, suppress printing of information on
prior distributions, regression coefficients and other parameters by
setting \code{priors}, \code{regcoef} and \code{otherpars} to
\code{FALSE}:

\begin{CodeChunk}

\begin{CodeInput}
R> list_models(mod2b, priors = FALSE, regcoef = FALSE, otherpars = FALSE)
\end{CodeInput}

\begin{CodeOutput}
Linear model for "SBP" 
   family: gaussian 
   link: identity 
* Predictor variables:
  (Intercept), ns(age, df = 2)1, ns(age, df = 2)2, 
  genderfemale, I(bili^2), I(bili^3) 


Linear model for "bili" 
   family: gaussian 
   link: identity 
* Predictor variables:
  (Intercept), age, genderfemale 
\end{CodeOutput}
\end{CodeChunk}

When a function of a variable is specified as auxiliary variable, this
function is used in the imputation models. For example, in \code{mod4b},
waist circumference (\code{WC}) is not part of the model for \code{SBP},
and the quadratic term \code{I(WC\textasciicircum 2)} is used in the
linear predictor of the imputation model for \code{bili}:

\begin{CodeChunk}

\begin{CodeInput}
R> mod4b <- lm_imp(SBP ~ age + gender + bili, auxvars = ~ I(WC^2),
+   data = NHANES)
R> 
R> list_models(mod4b, priors = FALSE, regcoef = FALSE, otherpars = FALSE)
\end{CodeInput}

\begin{CodeOutput}
Linear model for "SBP" 
   family: gaussian 
   link: identity 
* Predictor variables:
  (Intercept), age, genderfemale, bili 


Linear model for "bili" 
   family: gaussian 
   link: identity 
* Predictor variables:
  (Intercept), age, genderfemale, I(WC^2) 


Linear model for "WC" 
   family: gaussian 
   link: identity 
* Predictor variables:
  (Intercept), age, genderfemale 
\end{CodeOutput}
\end{CodeChunk}

Incomplete variables are always imputed on their original scale, i.e.,
in \code{mod2b} the variable \code{bili} is imputed and the quadratic
and cubic versions calculated from the imputed values. Likewise,
\code{creat} and \code{albu} in \code{mod2c} are imputed separately, and
\code{I(creat/albu\textasciicircum 2)} calculated from the imputed (and
observed) values. To ensure consistency between variables, functions
involving incomplete variables should be specified as part of the model
formula and not be pre-calculated as separate variables.

\hypertarget{sec:multilevel}{%
\subsection{Multi-level structure and longitudinal
covariates}\label{sec:multilevel}}

In multi-level models, additional to the fixed effects structure
specified by the argument \code{fixed}, a random effects structure needs
to be provided, either via the argument \code{random} (as in the package
\pkg{nlme}) or in round brackets (as in the package \pkg{lme4}).

The argument \code{random} takes a one-sided formula starting with a
\code{\textasciitilde}, and the grouping variable is separated by
\code{|}. A random intercept is added automatically and only needs to be
specified in a random intercept only model.

A few examples:

\begin{itemize}
\tightlist
\item
  random intercept only, with \code{id} as grouping variable:\newline
  \code{random = ~1 | id} or \code{formula = <...> + (1 | id)}
\item
  random intercept and slope for variable \code{time}:\newline
  \code{random = ~time | id} or \code{formula = <...> + (time  | id)}
\item
  random intercept, slope and quadratic random effect for
  \code{time}:\newline
  \code{random = ~time + I(time\textasciicircum 2) | id} or\newline
  \code{formula = <...> + (time + I(time\textasciicircum 2) | id)}
\item
  random intercept, random slope for \code{time} and random effect for
  variable \code{x}:\newline \code{random = ~time + x | id} or
  \code{formula = <...> + (time + x | id)}
\end{itemize}

It is possible to use splines in the random effects structure if there
are no missing values in the variables involved, e.g.:

\begin{CodeChunk}

\begin{CodeInput}
R> mod5 <- lme_imp(bmi ~ GESTBIR + ETHN + HEIGHT_M + ns(age, df = 2),
+   random = ~ ns(age, df = 2) | ID, data = simLong)
\end{CodeInput}
\end{CodeChunk}

To specify multiple levels of grouping, i.e., a hierarchical model with
more than two levels, the specification via the argument \code{formula}
should be used. Note that in \pkg{JointAI} there is no difference
between \code{(1 | id) +
(1 | center)} and \code{(1 | center/id)}. The distinction between nested
and crossed random effects needs to be done via the coding of the two
grouping variables: if \code{id} should be nested in \code{center} then
all cases with the same \code{id} have to have the same value for
\code{center}.

\hypertarget{sec:survmod}{%
\subsection{Survival models}\label{sec:survmod}}

\pkg{JointAI} provides two functions to analyse survival data with
incomplete covariates: \newline \code{survreg_imp()} and
\code{coxph_imp()}. Analogously to the complete data versions of these
functions from the package \pkg{survival}, the left hand side of the
model formula has to be a survival object specified using the function
\code{Surv()}.

\hypertarget{example-2}{%
\paragraph{Example:}\label{example-2}}

To analyse the \code{PBC} data, we can either use a parametric Weibull
model (considering only time-constant covariates) or a proportional
hazards model. Since the \code{PBC} data contains time-varying
covariates, we use the subset of rows where \code{day == 0} to use only
one observation per patient.

\begin{CodeChunk}

\begin{CodeInput}
R> mod6a <- survreg_imp(Surv(futime, status != "alive") ~ age + sex + 
R+   copper + trig, models = c(copper = "lognorm", trig = "lognorm"),
R+   data = subset(PBC, day == 0), n.iter = 250)
\end{CodeInput}
\end{CodeChunk}

\begin{CodeChunk}

\begin{CodeInput}
R> mod6b <- coxph_imp(Surv(futime, status != "alive") ~ age + sex + 
R+   copper + trig, models = c(copper = "lognorm", trig = "lognorm"),
R+   data = subset(PBC, day == 0), n.iter = 250)
\end{CodeInput}
\end{CodeChunk}

Currently only right-censored survival data can be handled and it is not
yet possible to take into account strata (i.e., strata specific baseline
hazards). To model clustered data, the model formula can be extended
with a random effects specification of the form
\code{formula = <...> + (1 | center)}. The specification of
subject-specific random effects also allows the user to include
time-varying covariates in proportional hazards models. This requires
the specification of the variable containing the timing of the
measurements of the time-varying measurements via the additional
argument \code{timevar}:

\begin{CodeChunk}

\begin{CodeInput}
R> mod6c <- coxph_imp(Surv(futime, status != "alive") ~ age + sex + copper +
R+   trig + platelet + (1 | id), 
R+   models = c(copper = "lognorm", trig = "lognorm"),
R+   timevar = "day", data = PBC)
\end{CodeInput}
\end{CodeChunk}

Time-varying covariates are modelled (and imputed) using the
last-observation-carried-forward principle. The data should include a
baseline measurement (where \code{timevar = 0}) of the time-varying
covariates. If a value needs to be filled in and no previous measurement
is available, the subsequent observation is ``carried-backward''.

\hypertarget{joint-models-1}{%
\subsection{Joint models}\label{joint-models-1}}

Joint models for longitudinal and survival data can be fitted using the
function \code{JM\_imp()}. The specification is analogue to the
specification of a proportional hazards model with time-dependent
covariates, but the longitudinal trajectories are assumed to follow a
smooth trajectory over time (as modelled by a mixed model) and not a
step-function.

If the models for time-varying covariates are not explicitly specified,
random intercept models with the default fixed effects structure are
used (including linear effects for all baseline variables and
time-varying variables that are complete or imputed earlier in the
sequence).

To specify models for time-dependent covariates, a list of models can be
supplied to the argument \code{formula}:

\begin{CodeChunk}

\begin{CodeInput}
R> PBC$logbili <- log(PBC$bili)
R> mod6d <- JM_imp(
R+   list(Surv(futime, status != "alive") ~ age + sex + platelet + logbili +
R+          (1 | id),
R+        platelet ~ age + sex + day + logbili + (day | id),
R+        logbili ~ age + sex + day + (day | id)
R+   ),
R+   timevar = "day", data = PBC, n.adapt = 10)
\end{CodeInput}
\end{CodeChunk}

The use of a list of model formulas is not restricted to \code{JM_imp()}
but possible in any of the main analysis functions. This allows the user
to fit multiple analyses simultaneously, or to explicitly specify the
structure of a covariate model. Joint analysis of multiple substantive
models may be particularly desirable if they share incomplete
covariates.

\hypertarget{sec:methJSS}{%
\subsection{Imputation / covariate model types}\label{sec:methJSS}}

\pkg{JointAI} automatically selects an (imputation) model type for each
of the incomplete covariates (and sometimes also complete covariates, as
detailed in Section~\ref{sec:impLong}) based on the \code{class} of the
variable.

The automatically selected types for covariates on the highest level
are:

\begin{itemize}
\tightlist
\item
  \code{lm}: linear model (for continuous variables)
\item
  \code{glm_binomial_logit}: binary logistic model (for factors with two
  levels)
\item
  \code{mlogit}: multinomial logit model (for unordered factors with
  \(>2\) levels)
\item
  \code{clm}: cumulative logit model (for ordered factors with \(>2\)
  levels)
\end{itemize}

The default methods for covariates on lower levels are:

\begin{itemize}
\tightlist
\item
  \code{lmm}: linear mixed model
\item
  \code{glmm_binomial_logit}: logistic mixed model
\item
  \code{mlogitmm}: multinomial logit mixed model
\item
  \code{clmm}: cumulative logit mixed model
\end{itemize}

When a continuous variable has only two different values, it is
automatically converted to a factor and modelled using a logistic model,
unless a different model type is specified by the user. Variables of
type \code{logical} are also converted to binary factors.

The (imputation) models that are chosen by default may not necessarily
be appropriate for the data at hand, especially for continuous
variables, which often do not comply with the assumptions of
(conditional) normality.

Therefore, the following alternative (imputation) model types are
available:

\begin{itemize}
\tightlist
\item
  Gamma (mixed) models for right-skewed variables \(>0\):\newline
  \code{glm\_gamma\_<link>} and \code{glmm\_gamma\_<link>}, where
  \code{<link>} should be one of \code{inverse}, \code{identity} or
  \code{log}
\item
  Poisson (mixed) models for count data: \newline
  \code{glm\_poisson\_<link>} and \code{glmm\_poisson\_<link>}, where
  \code{<link>} should be \code{log} or \code{identity}
\item
  Beta (mixed) models for for continuous variables with values in
  \((0, 1)\):\newline \code{beta} and \code{glmm\_beta}
\item
  Log-normal (mixed) model for right-skewed variables \(>0\):\newline
  \code{lognorm} and \code{glmm\_lognorm}
\end{itemize}

All model types are implemented as described in
Section~\ref{sec:AnalysisModel}

\hypertarget{specification-of-imputationcovariate-model-types}{%
\subsubsection{Specification of imputation/covariate model
types}\label{specification-of-imputationcovariate-model-types}}

In models \code{mod3b} and \code{mod3c} in Section~\ref{sec:restsupport}
we have already seen two examples in which the imputation model type was
changed using the argument \code{models}. This argument takes a named
vector of (imputation) model types, where the names are the names of
covariates. When the vector supplied to \code{models} only contains
specifications for a subset of the covariates for which a model is
needed, default models are used for the remaining ones. As explained in
Section~\ref{sec:impLong}, models for completely observed covariates may
need to be specified in multi-level settings.

\begin{CodeChunk}

\begin{CodeInput}
R> mod7a <- lm_imp(SBP ~ age + gender + WC + alc + bili + occup + smoke,
R+   models = c(WC = "glm_gamma_inverse", bili = "lognorm"), data = NHANES,
R+   n.adapt = 0)
R> 
R> mod7a$models
\end{CodeInput}

\begin{CodeOutput}
                    SBP                     alc                   occup 
"glm_gaussian_identity"    "glm_binomial_logit"                "mlogit" 
                   bili                   smoke                      WC 
              "lognorm"                   "clm"     "glm_gamma_inverse" 
\end{CodeOutput}
\end{CodeChunk}

When there is a ``time'' variable in the model, such as \code{age} in
our example (which is the age of the child at the time of the
measurement), it may not be meaningful to specify a model for that
variable. Especially when the ``time'' variable is pre-specified by the
design of the study it can usually be assumed to be independent of the
covariates and a model for it has no useful interpretation. The argument
\code{no_model} allows the user to exclude models for such variables (as
long as they are completely observed):

\begin{CodeChunk}

\begin{CodeInput}
R> mod7b <- lme_imp(bmi ~ GESTBIR + ETHN + HEIGHT_M + SMOKE + hc +
+   MARITAL + ns(age, df = 2), random = ~ ns(age, df = 2) | ID,
+   data = simLong, no_model = "age", n.adapt = 0)
R> 
R> mod7b$models
\end{CodeInput}

\begin{CodeOutput}
                     bmi                       hc                    SMOKE 
"glmm_gaussian_identity"                    "lmm"                    "clm" 
                 MARITAL                     ETHN                 HEIGHT_M 
                "mlogit"     "glm_binomial_logit"                     "lm" 
\end{CodeOutput}
\end{CodeChunk}

Note that by excluding the model for \code{age} we implicitly assume
that incomplete baseline variables are independent of \code{age}.

\hypertarget{order-of-the-sequence-of-imputation-models}{%
\subsubsection{Order of the sequence of imputation
models}\label{order-of-the-sequence-of-imputation-models}}

In multi-level models, the sequence of models for covariates is sorted
by the variables level, so that variables of a higher level enter the
linear predictor of variables of lower levels, but not vice versa.
Within each level, models are ordered by the number of missing values
(decreasing), so that the model for the variable with the most missing
values has the most variables in its linear predictor.

\hypertarget{sec:auxvars}{%
\subsection{Auxiliary variables}\label{sec:auxvars}}

Auxiliary variables are variables that are not part of the analysis
model but should be considered as predictor variables in the imputation
models because they can inform the imputation of unobserved values.

Good auxiliary variables are \citep{Buuren2012}

\begin{itemize}
\tightlist
\item
  associated with an incomplete variable of interest, or are
\item
  associated with the missingness of that variable and
\item
  do not have too many missing values themselves. Importantly, they
  should be observed for a large proportion of the cases that have a
  missing value in the variable to be imputed.
\end{itemize}

In the main functions \code{*_imp()}, auxiliary variables can be
specified with the argument \code{auxvars}, which takes a one-sided
formula.

\hypertarget{example-3}{%
\paragraph{Example:}\label{example-3}}

We might consider the variables \code{educ} and \code{smoke} as
predictors for the imputation of \code{occup}:

\begin{CodeChunk}

\begin{CodeInput}
R> mod8a <- lm_imp(SBP ~ gender + age + occup, auxvars = ~ educ + smoke,
+   data = NHANES, n.iter = 100)
\end{CodeInput}
\end{CodeChunk}

The variables \code{educ} and \code{smoke} are not included in the
analysis model. They are, however, used as predictors in the imputation
for \code{occup} and imputed themselves if they have missing values:

\begin{CodeChunk}

\begin{CodeInput}
R> list_models(mod8a, priors = FALSE, regcoef = FALSE, otherpars = FALSE,
+   refcat = FALSE)
\end{CodeInput}

\begin{CodeOutput}
Linear model for "SBP" 
   family: gaussian 
   link: identity 
* Predictor variables:
  (Intercept), genderfemale, age, occuplooking for work, occupnot working 


Multinomial logit model for "occup" 
* Predictor variables:
  (Intercept), genderfemale, age, educhigh, smokeformer, smokecurrent 


Cumulative logit model for "smoke" 
* Predictor variables:
  genderfemale, age, educhigh 
\end{CodeOutput}
\end{CodeChunk}

\hypertarget{functions-of-variables-as-auxiliary-variables}{%
\subsubsection{Functions of variables as auxiliary
variables}\label{functions-of-variables-as-auxiliary-variables}}

As shown above in \code{mod4b}, it is possible to specify functions of
auxiliary variables. In that case, the auxiliary variable is not
considered as a linear effect but as specified by the function.

Note that omitting auxiliary variables from the analysis model implies
that the outcome is independent of these variables, conditional on the
other variables in the model. If this is not true, the model is
misspecified which may lead to biased results (similar to leaving a
confounding variable out of a model).

\hypertarget{reference-values-for-categorical-covariates}{%
\subsection{Reference values for categorical
covariates}\label{reference-values-for-categorical-covariates}}

In \pkg{JointAI}, contrasts for incomplete categorical variables need to
be derived from the imputed values in each iteration of the MCMC
sampling. Currently, this is only implemented for dummy and effect
coding, i.e., \code{"contr.treatment}" and \code{"contr.sum"}. If a
model contains an incomplete ordered factor as covariate, and
\proglang{R}'s default \code{"contr.poly"} (orthogonal polynomials) for
ordered factors is set in the global \code{options()}, a warning is
printed and dummy coding is used instead.

By default, the first category of a categorical variable (ordered or
unordered) is used as reference, however, this may not always allow the
desired interpretation of the regression coefficients. Moreover, when
categories are unbalanced, setting the largest group as reference may
result in better mixing of the MCMC chains. Therefore, \pkg{JointAI}
allows the user to specify the reference category separately for each
variable, via the argument \code{refcats}. Changes in \code{refcats}
will not impact the imputation of the respective variable, but the
definition of the contrasts, which affects the linear predictor of the
analysis model or other covariate models.

\hypertarget{setting-reference-categories-for-all-variables}{%
\subsubsection{Setting reference categories for all
variables}\label{setting-reference-categories-for-all-variables}}

To specify the choice of reference category globally for all variables
in the model, \code{refcats} can be set as

\begin{itemize}
\item
  \code{refcats = "first"}
\item
  \code{refcats = "last"}
\item
  \code{refcats = "largest"}
\end{itemize}

For example:

\begin{CodeChunk}

\begin{CodeInput}
R> mod9a <- lm_imp(SBP ~ gender + age + race + educ + occup + smoke,
+   refcats = "largest", data = NHANES)
\end{CodeInput}
\end{CodeChunk}

\hypertarget{setting-reference-categories-for-individual-variables}{%
\subsubsection{Setting reference categories for individual
variables}\label{setting-reference-categories-for-individual-variables}}

Alternatively, \code{refcats} takes a named vector, in which the
reference category for each variable can be specified either by its
number or its name, or one of the three global types: \code{"first"},
\code{"last"} or \code{"largest"}. For variables for which no reference
category is specified in the list the default is used.

\begin{CodeChunk}

\begin{CodeInput}
R> mod9b <- lm_imp(SBP ~ gender + age + race + educ + occup + smoke,
+   refcats = list(occup = "not working", race = 3, educ = "largest"),
+   data = NHANES)
\end{CodeInput}
\end{CodeChunk}

To facilitate specification of the reference categories, the function
\code{set_refcat()} can be used. It prints the names of the categorical
variables that are selected by

\begin{itemize}
\tightlist
\item
  a specified model formula (using the argument \code{formula}) and/or
\item
  a one-sided formula specifying auxiliary variables (using the argument
  \code{auxvars}), or
\item
  a vector naming covariates (using the argument \code{covars})
\end{itemize}

or all categorical variables in the data if only the argument
\code{data} is provided, and asks a number of questions which the user
needs to reply to via input of a number.

\begin{CodeChunk}

\begin{CodeInput}
R> refs_mod9 <- set_refcat(NHANES, formula = formula(mod9b))
\end{CodeInput}

\begin{CodeOutput}
The categorical variables are:
- "gender"
- "race"
- "educ"
- "occup"
- "smoke"
\end{CodeOutput}

\begin{CodeOutput}

How do you want to specify the reference categories?

1: Use the first category for each variable.
2: Use the last category for each variabe.
3: Use the largest category for each variable.
4: Specify the reference categories individually.
\end{CodeOutput}
\end{CodeChunk}

When option 4 is chosen, a question for each categorical variable is
asked, for example:

\begin{CodeOutput}
The reference category for "race" should be 
 
1: Mexican American
2: Other Hispanic
3: Non-Hispanic White
4: Non-Hispanic Black
5: other
\end{CodeOutput}

After specification of the reference categories for all categorical
variables, the determined specification for the argument \code{refcats}
is printed:

\begin{CodeOutput}
In the JointAI model specify:
refcats = c(gender = "female", race = "Non-Hispanic White",
           educ = "low", occup = "not working", smoke = "never")

or use the output of this function.
\end{CodeOutput}

\code{set_refcat()} also returns a named vector that can be passed to
the argument \code{refcats}:

\begin{CodeChunk}

\begin{CodeInput}
R> mod9c <- lm_imp(SBP ~ gender + age + race + educ + occup + smoke,
+   refcats = refs_mod9, data = NHANES)
\end{CodeInput}
\end{CodeChunk}

\hypertarget{sec:hyperpars}{%
\subsection{Hyper-parameters}\label{sec:hyperpars}}

In a Bayesian framework, parameters are random variables for which a
distribution needs to be specified. These distributions depend on
parameters themselves, i.e., on hyper-parameters.

The function \code{default_hyperpars()} returns a list containing the
default hyper-parameters used in a \code{JointAI} model (see
Appendix~\ref{sec:AppHyperpars}).

\code{mu_reg_*} and \code{tau_reg_*} refer to the mean and precision of
the prior distribution for regression coefficients. \code{shape_tau_*}
and \code{rate_tau_*} are the shape and rate parameters of a gamma
distribution that is used as prior for precision parameters.
\code{RinvD} is the scale matrix in the Wishart prior for the inverse of
the random effects covariance matrix \code{D}, and \code{KinvD} is the
number of degrees of freedom in that distribution.
\code{shape_diag_RinvD} and \code{rate_diag_RinvD} are the shape and
rate parameters of the gamma prior of the diagonal elements of
\code{RinvD}. In random effects models with only one random effect, a
gamma prior is used instead of the Wishart distribution for the inverse
of \code{D}.

The hyper-parameters \code{mu_reg_surv} and \code{tau_reg_surv} are used
in \code{survreg_imp()} as well as \code{coxph_imp()} and
\code{JM_imp()}.

To change hyper-parameters in a \code{JointAI} model, the default values
can be obtained from \code{default_hyperpars()}, and then be adjusted
and passed to the argument \code{hyperpars}:

\begin{CodeChunk}

\begin{CodeInput}
R> hyp <- default_hyperpars()
R> hyp$norm["shape_tau_norm"] <- 0.5
R> 
R> mod9d <- lm_imp(SBP ~ gender + age + race + educ + occup + smoke,
+   data = NHANES, hyperpars = hyp)
\end{CodeInput}
\end{CodeChunk}

\vspace*{-3ex}

\hypertarget{scaling}{%
\subsection{Scaling}\label{scaling}}

When variables are measured on very different scales this can result in
slow convergence and bad mixing. Therefore, \pkg{JointAI} automatically
scales continuous variables to approximately have mean zero and standard
deviation one when they enter a linear predictor. Results are
transformed back to the original scale. To prevent scaling, the argument
\code{scale_vars} in \code{*_imp()} can be set to \code{FALSE}. When a
vector of term labels is supplied to \code{scale_vars}, only those terms
are scaled. By default, only the MCMC samples that are scaled back to
the scale of the data are stored in a \code{JointAI} object. When the
argument \code{keep_scaled_mcmc
= TRUE}, the scaled sample is also kept.

\hypertarget{shrinkage-priors}{%
\subsection{Shrinkage priors}\label{shrinkage-priors}}

Using the argument \code{shrinkage} it is possible to impose a penalty
on the regression coefficients of all or some sub-models. If
\code{shrinkage = "ridge"}, a ridge penalty is imposed on the regression
coefficients of all sub-models by specifying a
\(\text{Gamma}(0.01, 0.01)\) prior for the precision of the regression
coefficients instead of setting it to a fixed (small) value. It is also
possible to provide a named vector to \code{shrinkage}, where the names
should be the names of the response variables of models on which the
penalty should be imposed, and the values the type of shrinkage (e.g.
\code{shrinkage = c(SBP = "ridge")}).

\hypertarget{jags-model-file}{%
\subsection{JAGS model file}\label{jags-model-file}}

Using the user-specified or default settings described above,
\pkg{JointAI} writes the \proglang{JAGS} model. By default, the model is
written to a temporary file and deleted when the MCMC sampling has
finished. When the argument \code{keep_model} is set to \code{TRUE} the
model file will be kept. In any case, the \proglang{JAGS} model is
stored in the \code{JointAI} object as a character string. Arguments
\code{modelname} and \code{modeldir} allow the user to specify the name
of the file (including the ending, e.g., \code{.R} or \code{.txt}) and
the file location. When a file with that same name already exists in the
given location, a question is prompted giving the user the option to use
the existing file or to overwrite it. To prevent the question, the
argument \code{overwrite} can be set to \code{TRUE} or \code{FALSE}.

The functionality of using an existing \proglang{JAGS} model file
enables the user to make changes to the \proglang{JAGS} model that is
created automatically by \pkg{JointAI}, for example, to change the type
of prior distribution used for a particular parameter.

\pagebreak

\hypertarget{sec:MCMCSettings}{%
\section{MCMC settings}\label{sec:MCMCSettings}}

The main functions \code{*_imp()} have a number of arguments that
specify settings for the MCMC sampling:

\begin{itemize}
\tightlist
\item
  \code{n.chains}: number of MCMC chains
\item
  \code{n.adapt}: number of iterations in the adaptive phase
\item
  \code{n.iter}: number of iterations in the sampling phase
\item
  \code{thin}: thinning degree
\item
  \code{monitor_params}: parameters/nodes to be monitored
\item
  \code{seed}: optional seed value for reproducibility
\item
  \code{inits}: initial values
\end{itemize}

The first four, as well as the following two arguments, are passed
directly to functions from the \proglang{R} package \pkg{rjags}:

\begin{itemize}
\tightlist
\item
  \code{quiet}: should printing of information be suppressed?
\item
  \code{progress.bar}: type of progress bar (\code{"text"}, \code{"gui"}
  or \code{"none"})
\end{itemize}

In the following sections, the arguments listed above are explained in
more detail and examples are given.

\hypertarget{number-of-chains-iterations-and-samples}{%
\subsection{Number of chains, iterations and
samples}\label{number-of-chains-iterations-and-samples}}

\hypertarget{number-of-chains}{%
\subsubsection{Number of chains}\label{number-of-chains}}

To evaluate convergence of MCMC chains it is helpful to create multiple
chains that have different starting values. More information on how to
evaluate convergence and the specification of initial values can be
found in Sections~\ref{sec:grcrit} and \ref{sec:inits}, respectively.

The argument \code{n.chains} selects the number of chains (by default
\code{n.chains = 3}). For calculating the model summary, multiple chains
are merged.

\hypertarget{adaptive-phase}{%
\subsubsection{Adaptive phase}\label{adaptive-phase}}

\proglang{JAGS} has an adaptive mode, in which samplers are optimized
(for example the step size is adjusted). Samples obtained during the
adaptive mode do not form a Markov chain and are discarded. The argument
\code{n.adapt} controls the length of this adaptive phase.

The default value for \code{n.adapt} is 100, which works well in many of
the examples considered here. Complex models may require longer adaptive
phases. If the adaptive phase is not sufficient for \proglang{JAGS} to
optimize the samplers, a warning message will be printed (see example
below).

\hypertarget{sampling-iterations}{%
\subsubsection{Sampling iterations}\label{sampling-iterations}}

\code{n.iter} specifies the number of iterations in the sampling phase,
i.e., the length of the MCMC chain. How many samples are required to
reach convergence and to have sufficient precision (see also
Section~\ref{sec:mcerror}) depends on the complexity of data and model,
and may range from as few as 100 to several million.

\hypertarget{thinning}{%
\subsubsection{Thinning}\label{thinning}}

In settings with high autocorrelation, it may take many iterations
before a sample is created that sufficiently represents the whole range
of the posterior distribution. Processing of such long chains can be
slow and may cause memory issues. The parameter \code{thin} allows the
user to specify if and how much the MCMC chains should be thinned before
storing them. By default \code{thin =
1} is used, which corresponds to keeping all values. A value
\code{thin = 10} would result in keeping every 10th value and discarding
all other values.

\hypertarget{sec:MCMCexamples}{%
\paragraph{Example: default settings}\label{sec:MCMCexamples}}

Using the default settings \code{n.adapt = 100} and \code{thin = 1}, and
100 sampling iterations, a simple model would be

\begin{CodeChunk}

\begin{CodeInput}
R> mod10a <- lm_imp(SBP ~ alc, data = NHANES, n.iter = 100)
\end{CodeInput}
\end{CodeChunk}

The relevant part of the model summary (obtained with \code{summary()})
shows that the first 100 iterations (adaptive phase) were discarded, the
100 iterations that follow form the posterior sample, thinning was set
to ``1'', and that there are three chains.

\begin{CodeChunk}

\begin{CodeOutput}
[...]
MCMC settings:
Iterations = 101:200
Sample size per chain = 100 
Thinning interval = 1 
Number of chains = 3 
\end{CodeOutput}
\end{CodeChunk}

\hypertarget{example-insufficient-adaptation-phase}{%
\paragraph{Example: insufficient adaptation
phase}\label{example-insufficient-adaptation-phase}}

\begin{CodeChunk}

\begin{CodeInput}
R> mod10b <- lm_imp(SBP ~ alc, data = NHANES, n.adapt = 10, n.iter = 100)
\end{CodeInput}

\begin{CodeOutput}
Warning in rjags::jags.model(file = modelfile, data = data_list, inits
= inits, : Adaptation incomplete
\end{CodeOutput}

\begin{CodeOutput}
NOTE: Stopping adaptation
\end{CodeOutput}
\end{CodeChunk}

Specifying \code{n.adapt = 10} results in a warning message. The
relevant part of the model summary from the resulting model is:

\begin{CodeChunk}

\begin{CodeOutput}
[...]
MCMC settings:
Iterations = 11:110
Sample size per chain = 100 
Thinning interval = 1 
Number of chains = 3 
\end{CodeOutput}
\end{CodeChunk}

\hypertarget{example-thinning}{%
\paragraph{Example: thinning}\label{example-thinning}}

\begin{CodeChunk}

\begin{CodeInput}
R> mod10c <- lm_imp(SBP ~ alc, data = NHANES, n.iter = 500, thin = 10)
\end{CodeInput}
\end{CodeChunk}

Here, iterations 110 until 600 are used in the output, but due to a
thinning interval of ten, the resulting MCMC chains contain only 50
samples instead of 500, that is, the samples from iteration 110, 120,
130, \ldots

\begin{CodeChunk}

\begin{CodeOutput}
[...]
MCMC settings:
Iterations = 110:600
Sample size per chain = 50 
Thinning interval = 10 
Number of chains = 3 
\end{CodeOutput}
\end{CodeChunk}

\hypertarget{sec:monitorparams}{%
\subsection{Parameters to follow}\label{sec:monitorparams}}

Since \pkg{JointAI} uses \proglang{JAGS} \citep{JAGS} for performing the
MCMC sampling, and \proglang{JAGS} only saves the values of MCMC chains
for those nodes for which the user has specified that they should be
monitored, this is also the case in \pkg{JointAI}.

For this purpose, the main functions \code{*_imp()} have an argument
\code{monitor_params}, which takes a named list (or a named vector) with
possible entries given in Table~\ref{tab:monitorparams}. This table
contains a number of keywords that refer to (groups of) nodes. Each of
the keywords works as a switch and can be specified as \code{TRUE} or
\code{FALSE} (with the exception of \code{other}).

\begin{table}[!ht]
\label{tab:monitorparams}
\centering
\begin{tabular}{lp{0.8\linewidth}}
\toprule
name/key word & what is monitored\\
\midrule
\code{analysis_main} & \code{betas} and 
\code{sigma_main} (for models with a variance parameter),
\code{tau_main} (for beta models),
\code{gamma_main} (for cumulative logit models),
\code{shape_main} (for parametric survival models),
\code{D_main} (for multi-level models),
\code{basehaz} (for proportional hazards models)\\
\hspace{1em}\code{betas} & regression coefficients of the main model(s)\\
\hspace{1em}\code{tau_main} & precision of the residuals from the main model(s)\\
\hspace{1em}\code{sigma_main} & standard deviation of the residuals from the main model(s)\\
\addlinespace
\code{analysis_random} & \code{ranef_main}, \code{D_main},
    \code{invD_main}, \code{RinvD_main}\\
\hspace{1em}\code{ranef_main} & random effects of the main model(s)\\
\hspace{1em}\code{D_main} & covariance matrix of the random effects from the main model(s)\\
\hspace{1em}\code{invD_main} & inverse of \code{D_main}\\
\hspace{1em}\code{RinvD_main} & scale matrix in Wishart prior(s) for \code{invD_main}\\
\addlinespace
\code{other_models} & \code{alphas}, \code{tau_other}, \code{sigma_other},
    \code{gamma_other}, \code{delta_other}\\
\hspace{1em}\code{alphas} & regression coefficients in the covariate model(s)\\
\hspace{1em}\code{tau_other} & precision parameters of the residuals from covariate model(s)\\
\hspace{1em}\code{gamma_other} & intercepts in ordinal imputation models\\
\hspace{1em}\code{delta_other} & increments of ordinal intercepts\\
\addlinespace
\code{imps} & imputed values\\
\addlinespace
\code{ranef_other} & random effects of the covariate model(s)\\
\code{D_other} & covariance matrix of the random effects from the covariate model(s)\\
\code{invD_other} & inverse of \code{D_other}\\
\code{RinvD_other} & scale matrix in Wishart prior(s) for \code{invD_other}\\
\addlinespace
\code{other} & additional nodes\\
\bottomrule
\end{tabular}
\caption{Key words and names of (groups of) nodes that can be specified 
to be monitored using the argument \code{monitor\_params}.}
\end{table}

The default setting is \code{monitor_params = c(analysis_main = TRUE)},
i.e., only the main parameters of the analysis model are monitored, and
monitoring is switched off for all other parameters. To additionally
monitor the parameters of covariate models and imputed
values\linebreak \code{monitor_params =
c(other_models = TRUE, imps = TRUE)} would have to be specified.

It is possible to switch off sub-sets of the selected groups of nodes,
for example, to to monitor all random effects parameters of the main
model(s), but not the random effects themselves:
\code{monitor_params = c(analysis_random =
TRUE, ranef_main = FALSE)}.

The element \code{other} in \code{monitor_params} allows the
specification of one or multiple additional nodes to be monitored. When
\code{other} is used with more than one element, \code{monitor_params}
has to be a list. Here, as an example, we monitor the probability of
being in the \code{alc>=1} group for subjects one through three and the
expected value of the distribution of \code{creat} for the first
subject.

\begin{CodeChunk}

\begin{CodeInput}
R> mod11a <- lm_imp(SBP ~ gender + WC + alc + creat, data = NHANES, 
+   monitor_params = list(other = c("p_alc[1:3]", "mu_creat[1]")))
\end{CodeInput}
\end{CodeChunk}

Even though this example may not be particularly meaningful, in cases of
convergence issues it can be helpful to be able to monitor any node of
the model, not just the ones that are typically of interest.

More examples are given in the package vignettes.

\FloatBarrier

\hypertarget{sec:inits}{%
\subsection{Initial values}\label{sec:inits}}

Initial values are the starting point for the MCMC sampler. Setting good
initial values, i.e., values that are likely under the posterior
distribution, can speed up convergence. By default, the argument
\code{inits = NULL}, which means that initial values are generated
automatically by \proglang{JAGS}. It is also possible to supply initial
values directly as a list or as a function.

Initial values can be specified for every unobserved node, that is,
parameters and missing values, and it is possible to specify initial
values for only a subset of nodes.

When the initial values provided by the user do not have elements named
\code{".RNG.name"} or \code{".RNG.seed"}, \pkg{JointAI} will add those
elements, which specify the name and seed value of the random number
generator used for each chain. The argument \code{seed} allows the
specification of a seed value with which the starting values of the
random number generator, and, hence, the values of the MCMC sample can
be reproduced.

\pagebreak

\hypertarget{initial-values-in-a-list-of-lists}{%
\subsubsection{Initial values in a list of
lists}\label{initial-values-in-a-list-of-lists}}

A list of initial values should have the same length as the number of
chains, where each element is a named list of initial values and initial
values should differ between chains.

For example, to create initial values for the parameter vector
\code{beta} and the precision parameter \code{tau_SBP} in \code{mod11a}
from above for three chains the following syntax could be used:

\begin{CodeChunk}

\begin{CodeInput}
R> init_list <- lapply(1:3, function(i) {
+   list(beta = rnorm(5), 
+        tau_SBP = rgamma(1, 1, 1))
+ })
R> 
R> mod12a <- lm_imp(SBP ~ gender + WC + alc + creat, data = NHANES,
+   inits = init_list)
\end{CodeInput}
\end{CodeChunk}

The user provided lists of initial values (and starting values for the
random number generator) are stored in the \code{JointAI} object and can
be accessed via \code{mod11a$mcmc_settings$inits}.

\hypertarget{initial-values-as-a-function}{%
\subsubsection{Initial values as a
function}\label{initial-values-as-a-function}}

Initial values can be specified as a function. The function should
either take no arguments or a single argument called \code{chain}, and
return a named list that supplies values for one chain.

For example, to create initial values for the parameter vectors
\code{beta} and \code{alpha} in \code{mod11a}:

\begin{CodeChunk}

\begin{CodeInput}
R> inits_fun <- function() {
+   list(beta = rnorm(5),
+        alpha = rnorm(9))
+ }
R> 
R> mod12b <- lm_imp(SBP ~ gender + WC + alc + creat, data = NHANES,
+   inits = inits_fun)
\end{CodeInput}
\end{CodeChunk}

When a function is supplied, the function is evaluated by \pkg{JointAI}
and the resulting \code{list} is stored in the \code{JointAI} object.

\hypertarget{sec:NodesInits}{%
\subsubsection{For which nodes can initial values be
specified?}\label{sec:NodesInits}}

Initial values can be specified for all unobserved stochastic nodes,
i.e., parameters or unobserved data for which a distribution is
specified in the \proglang{JAGS} model. They have to be supplied in the
format of the parameter or unobserved value in the \proglang{JAGS}
model. To find out which nodes there are in a model and in which form
they have to be specified, the function \code{coef()} from package
\pkg{rjags} can be used to obtain a list with the current values of the
MCMC chains (by default the first chain) from a \proglang{JAGS} model
object. This object is contained in a \code{JointAI} object under the
name \code{model} (this requires at least one iteration in the adaptive
phase). Elements of the initial values should have the same structure as
the elements in this list of current values. For more details, see the
package vignettes.

\pagebreak

\hypertarget{parallel-sampling}{%
\subsection{Parallel sampling}\label{parallel-sampling}}

To reduce the computational time it is possible to perform sampling of
multiple MCMC chains in parallel. The packages \pkg{future}
\citep{future} and \pkg{doFuture} \citep{doFuture} can be used to
specify how parallel processes are handled. To specify that a model
should be run on four different workers, the following specification can
be used before fitting the model:

\begin{CodeChunk}

\begin{CodeInput}
R> libary("doFuture")
R> doFuture::registerDoFuture()
R> plan(multiprocess(workers = 4))
\end{CodeInput}
\end{CodeChunk}

This setting will remain for the entire \proglang{R} session, unless it
is explicitly re-set to sequential computation, for instance using the
following syntax:

\begin{CodeChunk}

\begin{CodeInput}
R> plan(sequential)
\end{CodeInput}
\end{CodeChunk}

\hypertarget{sec:Results}{%
\section{After fitting}\label{sec:Results}}

Any of the main functions \code{*_imp()} will return an object of class
\code{JointAI}. It contains the original data (\code{data}), information
on the type of model (\code{call}, \code{analysis_type}, \code{models},
\code{fixed}, \code{random}, \code{hyperpars}) and MCMC sampling
(\code{mcmc_settings}), the JAGS model (as object of class \code{jags}
in the element \code{model} and as string in the element
\code{jagsmodel}) and MCMC sample (\code{MCMC}; if a sample was
generated), information on the setting the model was run in
(\code{comp_info}; containing the start time, computational time,
\pkg{JointAI} version number, and, under the element \code{future},
information on the setting for parallel computation), and some
additional elements that are used by methods for objects of class
\code{JointAI} but are typically not of interest to the user.

In this section, we describe how the results from a \code{JointAI} model
can be visualized, summarized and evaluated. The functions described
here use, by default, the full MCMC sample and show only the parameters
of the analysis model. Arguments \code{start}, \code{end}, \code{thin}
and \code{exclude_chains} are available to select a subset of the
iterations of the MCMC sample that is used to calculate the summary. The
argument \code{subset} allows the user to control for which nodes the
summary or visualization is returned and follows the same logic as the
argument \code{monitor_params} in \code{*_imp()}. For \code{JointAI}
objects that include multiple main models (i.e, when a \code{list} of
formulas was supplied), the argument \code{outcome} can be used to
provide a vector of integers to select for which of the analysis models
the output should be shown. The use of these arguments is further
explained in Section~\ref{sec:subset}.

\hypertarget{visualizing-the-posterior-sample}{%
\subsection{Visualizing the posterior
sample}\label{visualizing-the-posterior-sample}}

The posterior sample can be visualized by two commonly used plots: a
trace plot, showing samples across iterations, and a plot of the
empirical density of the posterior sample.

\hypertarget{trace-plot}{%
\subsubsection{Trace plot}\label{trace-plot}}

A trace plot shows the sampled values per chain and node across
iterations. It allows the visual evaluation of convergence and mixing of
the chains and can be obtained with the function \code{traceplot()}.

\begin{CodeChunk}

\begin{CodeInput}
R> mod13a <- lm_imp(SBP ~ gender + WC + alc + creat, data = NHANES,
+   n.iter = 500, seed = 2020)
R> 
R> traceplot(mod13a)
\end{CodeInput}
\begin{figure}

{\centering \includegraphics[width=0.9\linewidth]{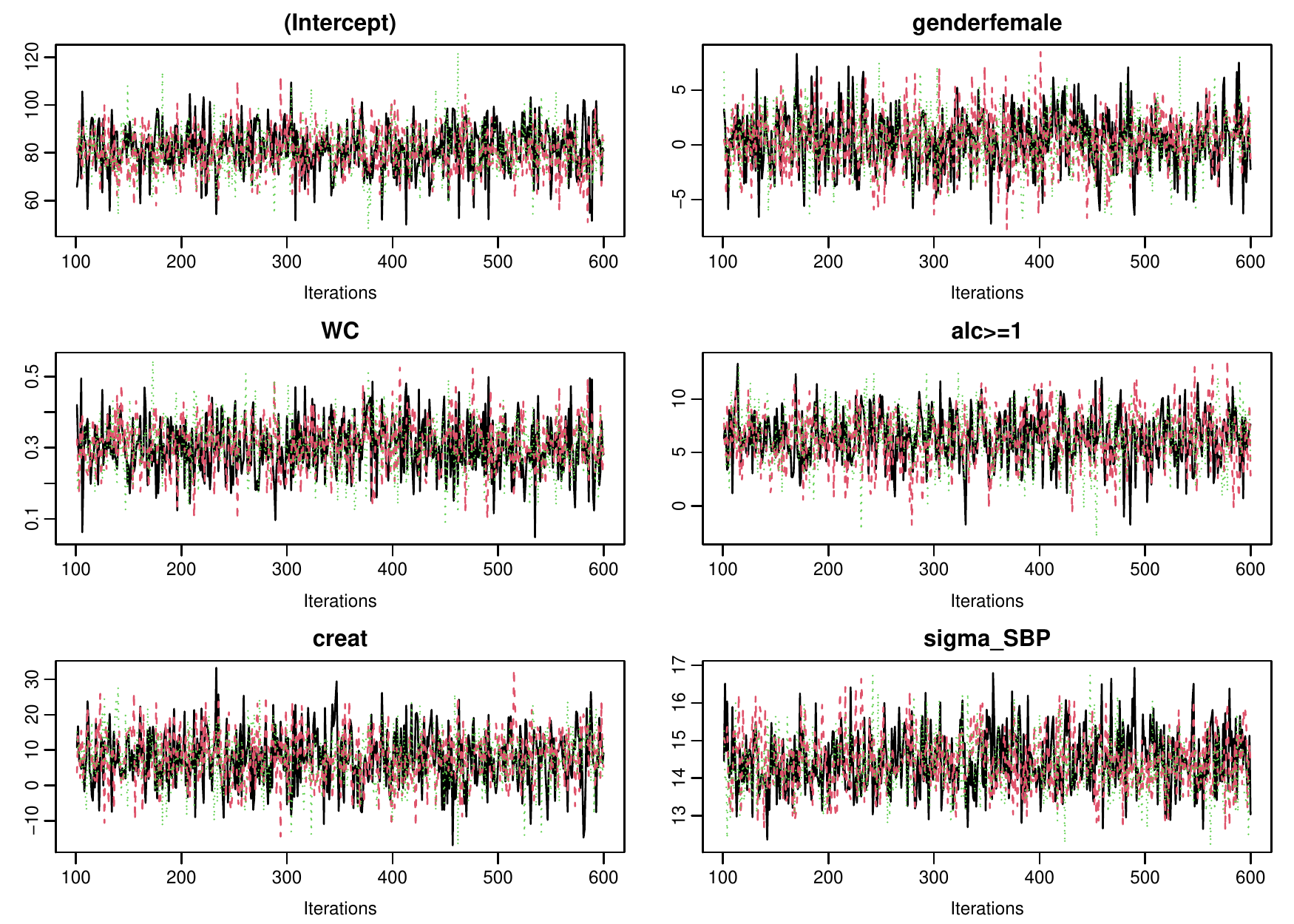} 

}

\caption[Traceplot of \code{mod13a}]{Traceplot of \code{mod13a}.}\label{fig:traceplot13a}
\end{figure}
\end{CodeChunk}

When the sampler has converged the chains show one horizontal band, as
in Figure~\ref{fig:traceplot13a}. Consequently, when traces show a
trend, convergence has not been reached and more iterations are
necessary (e.g., using \code{add_samples()}).

Graphical aspects of the trace plot can be controlled by specifying
standard graphical arguments via the dots argument \code{"..."}, which
are passed to \code{matplot()} (which is part of base \proglang{R}).
This allows the user to change colour, line type and -width, limits, and
so on. Arguments \code{nrow} and/or \code{ncol} can be supplied to set
specific numbers of rows and columns for the layout of the grid of
plots.

With the argument \code{use_ggplot} it is possible to get a
\pkg{ggplot2} \citep{ggplot2} version of the trace plot. It can be
extended using standard \pkg{ggplot2} syntax. The output of the
following syntax is shown in Figure~\ref{fig:ggtrace13a}.

\begin{CodeChunk}

\begin{CodeInput}
R> library("ggplot2")
R> traceplot(mod13a, ncol = 4, use_ggplot = TRUE) +
+   theme(legend.position = "bottom") +
+   scale_color_viridis_d(end = 0.9)
\end{CodeInput}
\begin{figure}

{\centering \includegraphics[width=1\linewidth]{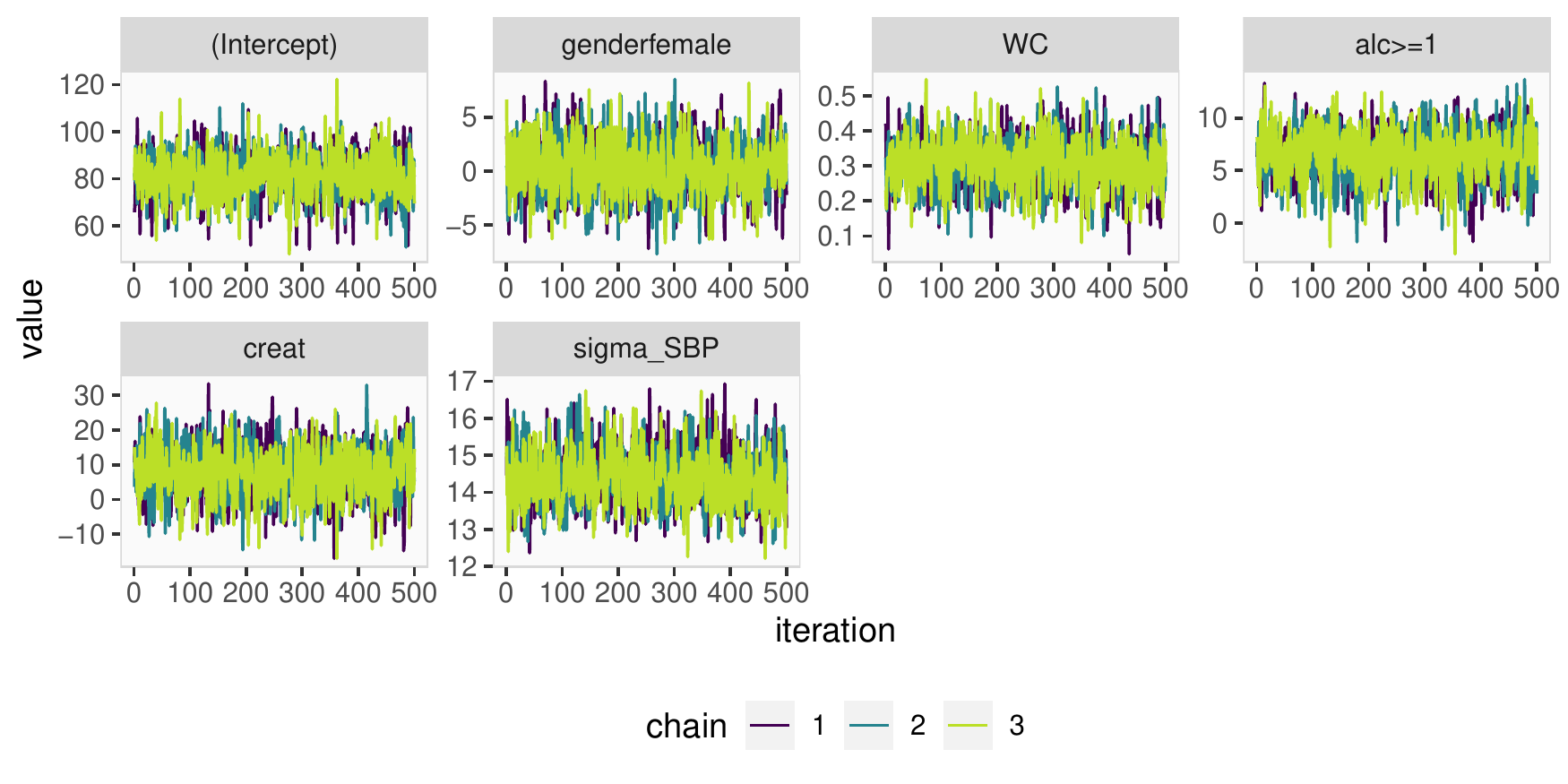} 

}

\caption[\pkg{ggplot2} version of the trace plot for \code{mod13a}]{\pkg{ggplot2} version of the trace plot for \code{mod13a}}\label{fig:ggtrace13a}
\end{figure}
\end{CodeChunk}

\hypertarget{density-plot}{%
\subsubsection{Density plot}\label{density-plot}}

The posterior distributions can also be visualized using the function
\code{densplot()}, which plots the empirical density per node per chain,
or combining chains (when \code{joined = TRUE}).

\begin{CodeChunk}

\begin{CodeInput}
R> densplot(mod13a, ncol = 4,
+   vlines = list(list(v = summary(mod13a)$stats[, "Mean"], lwd = 2),
+                 list(v = summary(mod13a)$stats[, "2.5%"], lty = 2),
+                 list(v = summary(mod13a)$stats[, "97.5%"], lty = 2)
+          )
+ )
\end{CodeInput}
\begin{figure}

{\centering \includegraphics[width=1\linewidth]{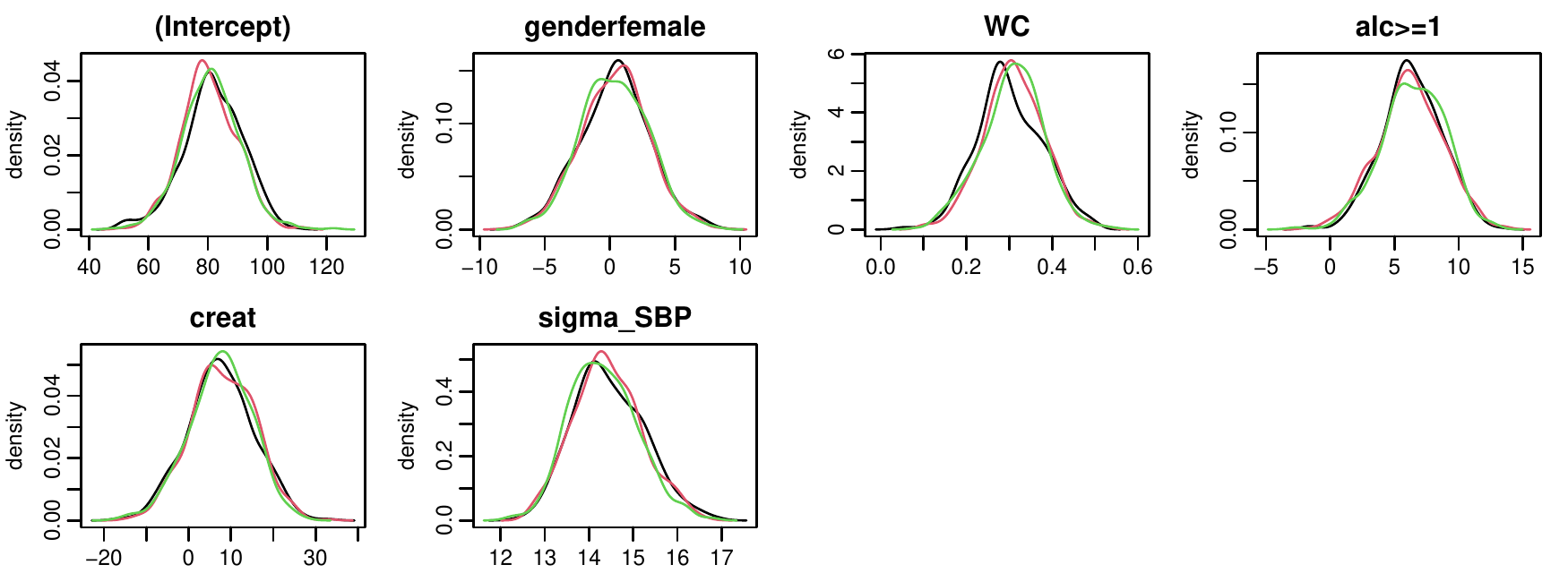} 

}

\caption[Empirical posterior density of \code{mod13a}]{Empirical posterior density of \code{mod13a}}\label{fig:densplot13a}
\end{figure}
\end{CodeChunk}

The argument \code{vlines} takes a list of lists, containing
specifications passed to the function \code{abline()} (part of base
\proglang{R}), and allows the addition of (vertical) lines to the plot,
e.g., marking zero, or marking the posterior mean and 2.5\% and 97.5\%
quantiles (Figure~\ref{fig:densplot13a}).

As with \code{traceplot()}, it is possible to use the \pkg{ggplot2}
version of \code{densplot()} when setting \code{use_ggplot = TRUE}.
Here, vertical lines can be added as additional layers.
Figure~\ref{fig:ggdens13a} shows, as an example, the posterior density
from \code{mod13a} to which vertical lines, representing the 95\%
credible interval and a 95\% confidence interval from a complete case
analysis, are added. The corresponding syntax is given in
Appendix~\ref{sec:AppDensplot}.

\begin{CodeChunk}
\begin{figure}

{\centering \includegraphics[width=1\linewidth]{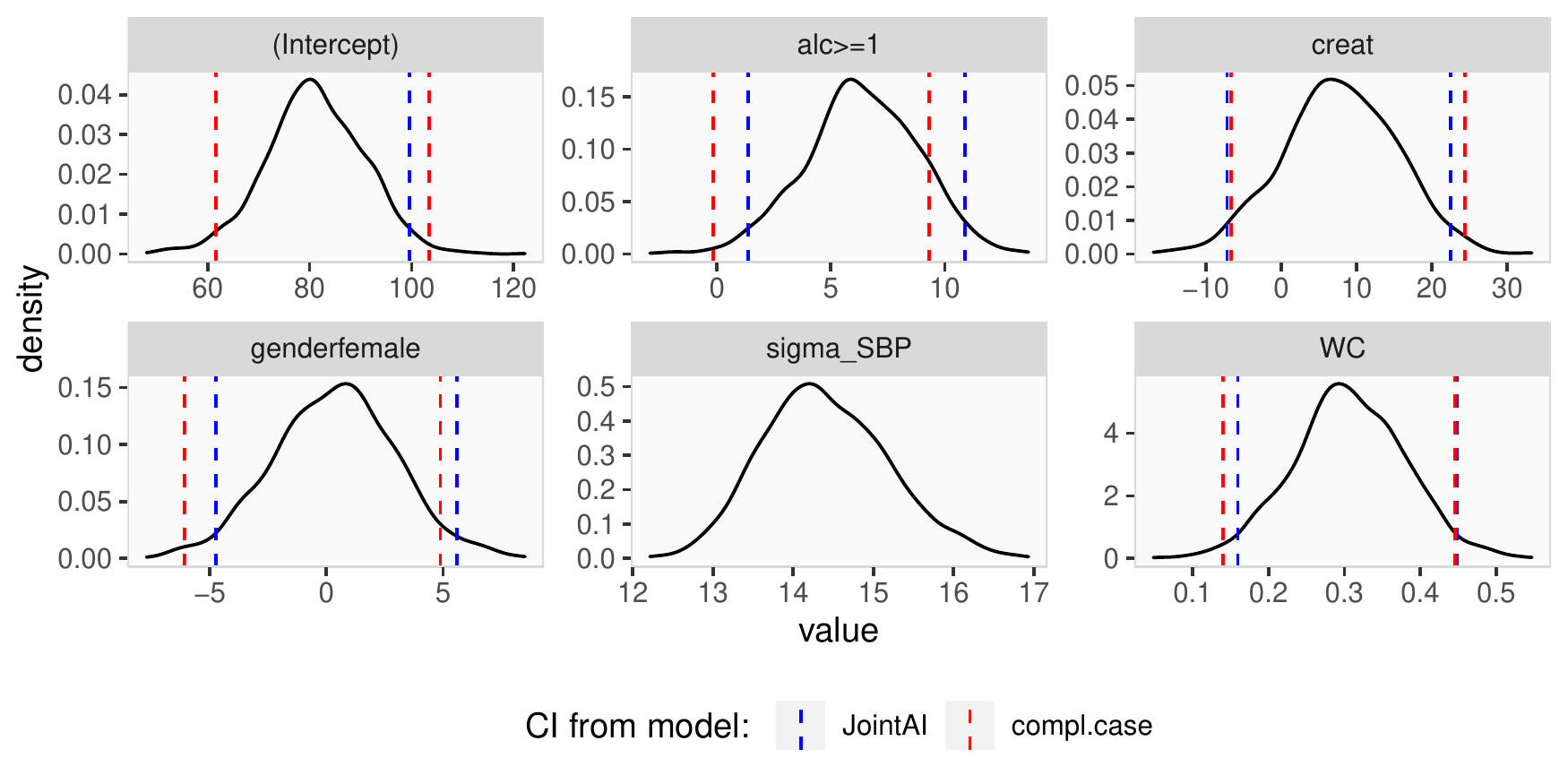} 

}

\caption[Density plot of model \code{mod13a}]{Density plot of model \code{mod13a}}\label{fig:ggdens13a}
\end{figure}
\end{CodeChunk}

\FloatBarrier

\hypertarget{model-summary}{%
\subsection{Model Summary}\label{model-summary}}

A summary of the posterior distribution estimated in a \code{JointAI}
model can be obtained using the function \code{summary()}.

The posterior summary consists of the mean, standard deviation and
quantiles (by default the 2.5\% and 97.5\% quantiles) of the MCMC
samples from all chains combined, as well as the tail probability (see
below), Gelman-Rubin criterion (see Section~\ref{sec:grcrit}) and Monte
Carlo error to posterior standard deviation ratio (see
Section~\ref{sec:mcerror}).

Additionally, some important characteristics of the MCMC samples on
which the summary is based, are given. This includes the range and
number of iterations (\code{Sample size per chain}), thinning interval
and number of chains. Furthermore, the number of observations (number of
rows in the data) is printed.

\begin{CodeChunk}

\begin{CodeInput}
R> summary(mod13a)
\end{CodeInput}

\begin{CodeOutput}

Bayesian linear model fitted with JointAI

Call:
lm_imp(formula = SBP ~ gender + WC + alc + creat, data = NHANES, 
    n.iter = 500, seed = 2020)


Posterior summary:
               Mean     SD  2.5%  97.5% tail-prob. GR-crit MCE/SD
(Intercept)  81.077 9.6921 61.66 99.602      0.000   1.011 0.0261
genderfemale  0.368 2.6138 -4.74  5.594      0.871   0.999 0.0258
WC            0.306 0.0736  0.16  0.448      0.000   1.012 0.0259
alc>=1        6.365 2.4692  1.38 10.897      0.016   1.006 0.0320
creat         7.747 7.5949 -7.19 22.496      0.299   1.003 0.0313

Posterior summary of residual std. deviation:
          Mean    SD 2.5% 97.5% GR-crit MCE/SD
sigma_SBP 14.4 0.779   13    16    1.02 0.0297


MCMC settings:
Iterations = 101:600
Sample size per chain = 500 
Thinning interval = 1 
Number of chains = 3 

Number of observations: 186 
\end{CodeOutput}
\end{CodeChunk}

Depending on the type of model, the output shows additional sections
with posterior summaries for model specific parameters, for example, the
random effects variance-covariance matrix for multi-level models or the
shape parameter of the Weibull distribution in a parametric survival
model. Using the argument \code{missinfo} information on the number and
proportion of complete cases and missing values per variable can be
added:

\begin{CodeChunk}

\begin{CodeInput}
R> library("splines")
R> mod13b <- lme_imp(bmi ~ GESTBIR + ETHN + HEIGHT_M + ns(age, df = 3),
+   random = ~ ns(age, df = 3) | ID, data = subset(simLong, !is.na(bmi)),
+   n.iter = 250)
R> 
R> summary(mod13b, missinfo = TRUE)
\end{CodeInput}

\begin{CodeOutput}

Bayesian linear mixed model fitted with JointAI

Call:
lme_imp(fixed = bmi ~ GESTBIR + ETHN + HEIGHT_M + ns(age, df = 3), 
    data = subset(simLong, !is.na(bmi)), random = ~ns(age, df = 3) | 
        ID, n.iter = 250)


Posterior summary:
                     Mean      SD    2.5%   97.5% tail-prob. GR-crit MCE/SD
(Intercept)      16.19417 2.28164 11.5856 20.4586      0.000    1.05 0.1069
GESTBIR          -0.03171 0.04501 -0.1137  0.0542      0.504    1.00 0.0678
ETHNother         0.02148 0.14415 -0.2752  0.2875      0.819    1.05 0.1237
HEIGHT_M          0.00542 0.00932 -0.0139  0.0247      0.539    1.08 0.1008
ns(age, df = 3)1 -0.28810 0.08061 -0.4181 -0.1267      0.000    3.57 0.7217
ns(age, df = 3)2  1.71880 0.13696  1.4451  1.9797      0.000    1.95 0.3488
ns(age, df = 3)3 -1.27190 0.04081 -1.3320 -1.1670      0.000    2.40 0.3335

Posterior summary of random effects covariance matrix:
                Mean     SD   2.5%  97.5% tail-prob. GR-crit MCE/SD
D_bmi_ID[1,1]  1.465 0.1733  1.168  1.847               1.01 0.0519
D_bmi_ID[1,2] -0.759 0.1118 -0.990 -0.560          0    1.05 0.0739
D_bmi_ID[2,2]  0.691 0.1085  0.518  0.929               1.29 0.2592
D_bmi_ID[1,3] -2.581 0.3672 -3.345 -1.953          0    1.01 0.0560
D_bmi_ID[2,3]  2.347 0.2822  1.867  2.922          0    1.05 0.0478
D_bmi_ID[3,3]  8.291 0.9848  6.622 10.339               1.02 0.0471
D_bmi_ID[1,4] -0.686 0.0958 -0.886 -0.522          0    1.06 0.0656
D_bmi_ID[2,4]  0.578 0.0737  0.447  0.721          0    1.06 0.1382
D_bmi_ID[3,4]  2.005 0.2500  1.554  2.519          0    1.26 0.1309
D_bmi_ID[4,4]  0.501 0.0731  0.370  0.654               1.55 0.2183

Posterior summary of residual std. deviation:
           Mean      SD  2.5% 97.5% GR-crit MCE/SD
sigma_bmi 0.458 0.00813 0.443 0.474    1.01 0.0495


MCMC settings:
Iterations = 101:350
Sample size per chain = 250 
Thinning interval = 1 
Number of chains = 3 

Number of observations: 1881 
Number of groups:
 - ID: 200


Number and proportion of complete cases:
        level    #   %
ID         ID  190  95
lvlone lvlone 1881 100

Number and proportion of missing values:
     level # NA % NA
bmi lvlone    0    0
age lvlone    0    0

         level # NA % NA
GESTBIR     ID    0    0
ID          ID    0    0
HEIGHT_M    ID    4    2
ETHN        ID    6    3
\end{CodeOutput}
\end{CodeChunk}

\FloatBarrier

\hypertarget{tail-probability}{%
\subsubsection{Tail probability}\label{tail-probability}}

The tail probability, calculated as
\(2\times\min\left\{Pr(\theta > 0), Pr(\theta < 0)\right\},\) where
\(\theta\) is the parameter of interest, is a measure of how likely the
value 0 is under the estimated posterior distribution.
Figure~\ref{fig:tailprob} visualizes three examples of posterior
distributions and the corresponding minimum of \(Pr(\theta > 0)\) and
\(Pr(\theta < 0)\) (shaded area).

\begin{CodeChunk}
\begin{figure}

{\centering \includegraphics[width=1\linewidth]{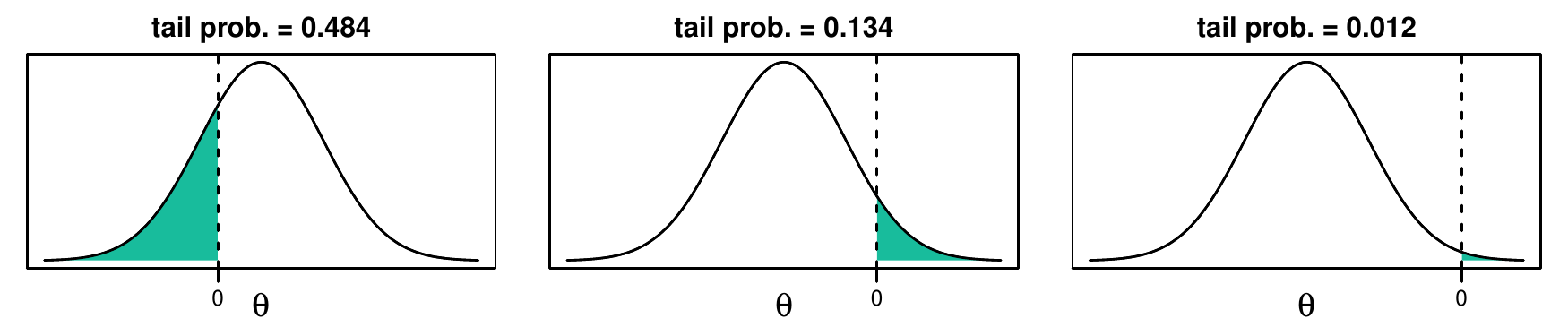} 

}

\caption[Visualization of the tail probability]{Visualization of the tail probability.}\label{fig:tailprob}
\end{figure}
\end{CodeChunk}

\FloatBarrier

\hypertarget{evaluation-criteria}{%
\subsection{Evaluation criteria}\label{evaluation-criteria}}

Convergence of the MCMC chains and precision of the posterior sample can
also be evaluated in a more formal manner. Implemented in \pkg{JointAI}
are the Gelman-Rubin criterion for convergence
\citep{Gelman1992, Brooks1998} and a comparison of the Monte Carlo error
with the posterior standard deviation.

\hypertarget{sec:grcrit}{%
\subsubsection{Gelman-Rubin criterion for
convergence}\label{sec:grcrit}}

The Gelman-Rubin criterion \citep{Gelman1992, Brooks1998}, also referred
to as ``potential scale reduction factor'', evaluates convergence by
comparing within and between chain variability and, thus, requires at
least two MCMC chains to be calculated. It is implemented for
\code{JointAI} objects in the function \code{GR_crit()}, which is based
on the function \code{gelman.diag()} from the package \pkg{coda}
\citep{coda}. The upper limit of the confidence interval should not be
much larger than 1.

\begin{CodeChunk}

\begin{CodeInput}
R> GR_crit(mod13a)
\end{CodeInput}

\begin{CodeOutput}
Potential scale reduction factors:

             Point est. Upper C.I.
(Intercept)        1.01       1.02
genderfemale       1.00       1.01
WC                 1.00       1.01
alc>=1             1.00       1.01
creat              1.00       1.00
sigma_SBP          1.02       1.05

Multivariate psrf

1.01
\end{CodeOutput}
\end{CodeChunk}

Besides the arguments \code{start}, \code{end}, \code{thin},
\code{exclude_chains} and \code{subset} (explained in
Section~\ref{sec:subset}) \code{GR_crit()} also takes the arguments
\code{confidence}, \code{transform} and \code{autoburnin} of
\code{gelman.diag()}.

\hypertarget{sec:mcerror}{%
\subsubsection{Monte Carlo Error}\label{sec:mcerror}}

Precision of the MCMC sample can be checked with the function
\code{MC_error()}. It uses the function \code{mcse()} from the package
\pkg{mcmcse} \citep{mcmcse} to calculate the Monte Carlo error (the
error that is made since the sample is finite) and compares it to the
standard deviation of the posterior sample. A rule of thumb is that the
Monte Carlo error should not be more than 5\% of the standard deviation
\citep{Lesaffre2012}. Besides the arguments explained in
Section~\ref{sec:subset}, \code{MC_error()} takes the arguments of
\code{mcse()}.

\begin{CodeChunk}

\begin{CodeInput}
R> MC_error(mod13a)
\end{CodeInput}

\begin{CodeOutput}
               est   MCSE    SD MCSE/SD
(Intercept)  81.08 0.2533 9.692   0.026
genderfemale  0.37 0.0675 2.614   0.026
WC            0.31 0.0019 0.074   0.026
alc>=1        6.37 0.0790 2.469   0.032
creat         7.75 0.2380 7.595   0.031
sigma_SBP    14.40 0.0232 0.779   0.030
\end{CodeOutput}
\end{CodeChunk}

\code{MC_error()} returns an object of class \code{MCElist}, which is a
list containing matrices with the posterior mean, estimated Monte Carlo
error, posterior standard deviation and the ratio of the Monte Carlo
error and posterior standard deviation, for the scaled (if this MCMC
sample was included in the \code{JointAI} object) and unscaled
(transformed back to the scale of the data) posterior samples. The
associated print method prints only the latter.

To facilitate quick evaluation of the Monte Carlo error to posterior
standard deviation ratio, plotting of an object of class \code{MCElist}
using \code{plot()} shows this ratio for each (selected) node and
automatically adds a vertical line at the desired cut-off (by default
5\%; see Figure~\ref{fig:MCE13a}):

\begin{CodeChunk}

\begin{CodeInput}
R> plot(MC_error(mod13a))  
R> plot(MC_error(mod13a, end = 250))
\end{CodeInput}
\end{CodeChunk}

\begin{CodeChunk}
\begin{figure}

{\centering \includegraphics[width=1\linewidth]{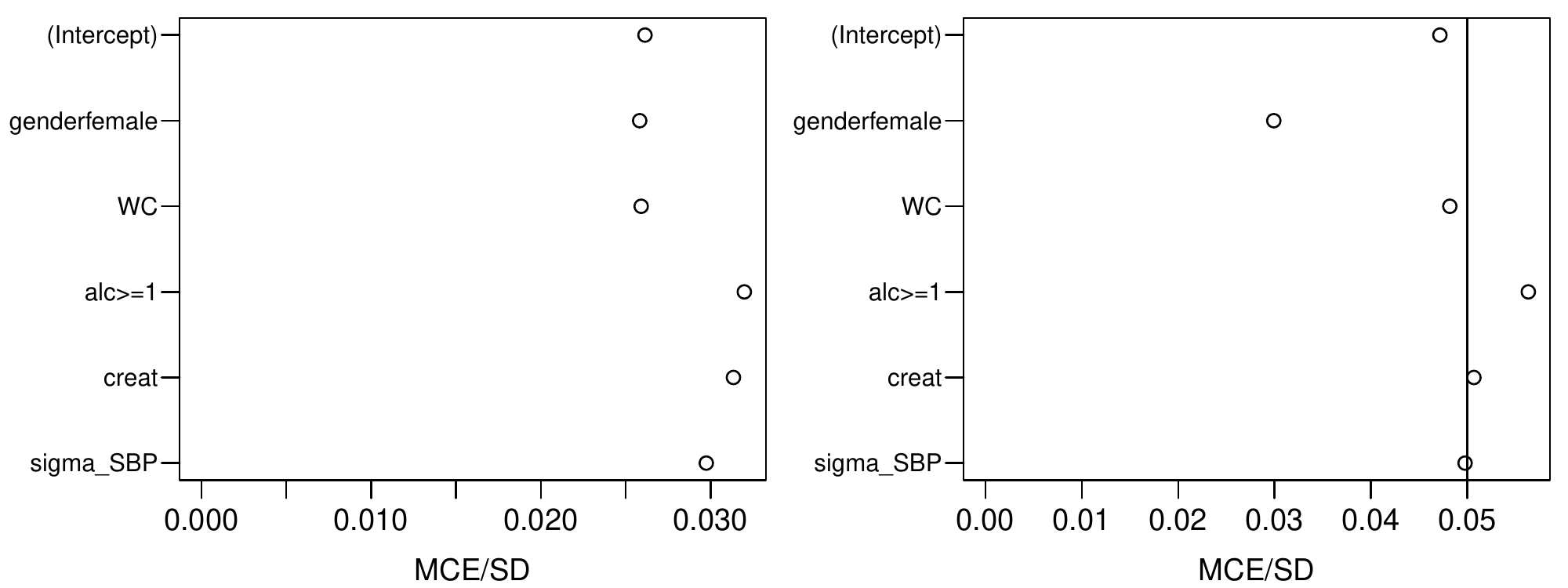} 

}

\caption[Plot of the MCElist object from \code{mod13a}]{Plot of the MCElist object from \code{mod13a}. Left: including
all iterations, right: using only the first 250 iterations of the MCMC sample.}\label{fig:MCE13a}
\end{figure}
\end{CodeChunk}

\hypertarget{sec:subset}{%
\subsection{Subset of the MCMC sample}\label{sec:subset}}

By default, the functions \code{traceplot()}, \code{densplot()},
\code{summary()}, \code{predict()}, \code{GR_crit()} and
\code{MC_error()} use all iterations of the MCMC sample and consider
only the parameters of the analysis model (if they were monitored). In
this section we describe how the set of iterations and parameters to
display can be changed using the arguments \code{subset}, \code{start},
\code{end}, \code{thin} and \code{exclude_chains}.

\hypertarget{subset-of-parameters}{%
\subsubsection{Subset of parameters}\label{subset-of-parameters}}

When the main parameters of the main/analysis model(s) have been
monitored in a \code{JointAI} object only these parameters are returned
in the model summary, plots and criteria shown above. If the main
parameters of the analysis model(s) were not monitored and the argument
\code{subset} is not specified, all parameters that were monitored are
displayed.

To display output for nodes other than the main parameters of the
analysis model or for a subset of nodes, the argument \code{subset}
needs to be specified. It follows the same logic as the argument
\code{monitor_params} of \code{*_imp} explained in
Section~\ref{sec:monitorparams}.

\hypertarget{example-4}{%
\paragraph{Example:}\label{example-4}}

To display only the parameters of the covariate models, we re-estimate
the model with the monitoring for these parameters switched on ans set
\code{subset = c(analysis_main = FALSE, other_models = TRUE)}:

\begin{CodeChunk}

\begin{CodeInput}
R> mod13c <- update(mod13a, monitor_params = c(other_models = TRUE))
R> summary(mod13c, subset = c(analysis_main = FALSE, other_models = TRUE))
\end{CodeInput}

\begin{CodeOutput}

Bayesian joint model fitted with JointAI 

Call:
lm_imp(formula = SBP ~ gender + WC + alc + creat, data = NHANES, 
    n.iter = 500, monitor_params = c(other_models = TRUE), seed = 2020)


# --------------------------------------------------------------------- #
  Bayesian binomial model for "alc"
# - - - - - - - - - - - - - - - - - - - - - - - - - - - - - - - - - - - #

Posterior summary:
                 Mean     SD    2.5%   97.5% tail-prob. GR-crit MCE/SD
(Intercept)   0.51390 1.5325 -2.6068  3.4078     0.7080    1.01 0.0567
genderfemale -0.88236 0.3995 -1.6322 -0.0498     0.0373    1.04 0.0738
WC            0.00632 0.0115 -0.0169  0.0298     0.5627    1.01 0.0375
creat        -1.48151 1.2238 -3.9056  0.9785     0.2213    1.00 0.0584


# --------------------------------------------------------------------- #
  Bayesian linear model for "creat"
# - - - - - - - - - - - - - - - - - - - - - - - - - - - - - - - - - - - #

Posterior summary:
                  Mean       SD      2.5%    97.5% tail-prob. GR-crit MCE/SD
(Intercept)   0.844704 0.076409  0.694938  0.99127      0.000       1 0.0258
genderfemale -0.178815 0.022122 -0.223627 -0.13699      0.000       1 0.0258
WC            0.000877 0.000772 -0.000612  0.00243      0.256       1 0.0258

Posterior summary of residual std. deviation:
             Mean      SD  2.5% 97.5% GR-crit MCE/SD
sigma_creat 0.145 0.00769 0.132 0.161    1.01       


# --------------------------------------------------------------------- #
  Bayesian linear model for "WC"
# - - - - - - - - - - - - - - - - - - - - - - - - - - - - - - - - - - - #

Posterior summary:
              Mean   SD  2.5%   97.5% tail-prob. GR-crit MCE/SD
(Intercept)  97.41 1.52 94.48 100.469     0.0000    1.00 0.0258
genderfemale -5.16 2.21 -9.38  -0.971     0.0147    1.01 0.0258

Posterior summary of residual std. deviation:
         Mean    SD 2.5% 97.5% GR-crit MCE/SD
sigma_WC 14.5 0.785 13.2  16.3       1 0.0258


# ----------------------------------------------------------- #

MCMC settings:
Iterations = 101:600
Sample size per chain = 500 
Thinning interval = 1 
Number of chains = 3 

Number of observations: 186 
\end{CodeOutput}
\end{CodeChunk}

\hypertarget{example-5}{%
\paragraph{Example:}\label{example-5}}

To select only some of the parameters, they can be specified directly by
name via the \code{other} element of \code{subset} (output not shown).

\begin{CodeChunk}

\begin{CodeInput}
R> densplot(mod13a, nrow = 1,
+   subset = list(analysis_main = FALSE, other = c("beta[2]", "beta[4]")))
\end{CodeInput}
\end{CodeChunk}

The function \code{parameters()} returns a \code{data.frame} containing
the names of all noes monitored in a \code{JointAI object} and can help
to identify the correct names of the nodes to be plotted.

\hypertarget{example-6}{%
\paragraph{Example:}\label{example-6}}

This also works when a subset of the imputed values should be displayed.

Re-fit the model and monitor the imputed values and select all imputed
values for \code{WC} (4th column of \code{M_lvlone}, the data matrix
containing all level-1 variables):

\begin{CodeChunk}

\begin{CodeInput}
R> mod13d <- update(mod13a, monitor_params = c(imps = TRUE))
\end{CodeInput}
\end{CodeChunk}

\begin{CodeChunk}

\begin{CodeInput}
R> sub3 <- grep("M_lvlone\\[[[:digit:]]+,4\\]", parameters(mod13d)$coef,
+   value = TRUE)
R> sub3
\end{CodeInput}

\begin{CodeOutput}
[1] "M_lvlone[33,4]"  "M_lvlone[150,4]"
\end{CodeOutput}
\end{CodeChunk}

Pass \code{sub3} to \code{subset} via \code{"other"}, for example in a
\code{traceplot()}.

\hypertarget{example-7}{%
\paragraph{Example:}\label{example-7}}

When the number of imputed values is large or in order to check
convergence of random effects, it may not be feasible to plot and
inspect all trace plots. In that case, a random subset of, for instance,
the random effects, can be selected.

Re-fit the model monitoring the random effects, then obtain a vector
with the names of all random effects and plot trace plots for a random
subset (output not shown):

\begin{CodeChunk}

\begin{CodeInput}
R> mod13e <- update(mod13b, monitor_params = c(ranef_main = TRUE))
\end{CodeInput}
\end{CodeChunk}

\begin{CodeChunk}

\begin{CodeInput}
R> rde <- grep("^b_bmi_ID\\[", colnames(mod13e$MCMC[[1]]), value = TRUE)
R> traceplot(mod13e, subset = list(analysis_main = FALSE,
+                                 other = sample(rde, size = 12)), ncol = 4)
\end{CodeInput}
\end{CodeChunk}

\hypertarget{sec:startend}{%
\subsubsection{Subset of MCMC samples}\label{sec:startend}}

With the arguments \code{start}, \code{end} and \code{thin} it is
possible to select which iterations from the MCMC sample are included in
the summary. \code{start} and \code{end} specify the first and last
iterations to be used, \code{thin} the thinning interval. Specification
of \code{start} thus allows the user to discard a ``burn-in'', i.e., the
iterations before the MCMC chain had converged.

If a particular chain has not converged it can be excluded from the
result summary or plot using the argument \code{exclude_chains} which
takes a numeric vector identifying chains to be excluded, e.g.,
\code{exclude_chains = c(1, 3)}.

\hypertarget{predicted-values}{%
\subsection{Predicted values}\label{predicted-values}}

Often, the aim of an analysis is not only to estimate the association
between outcome and covariates but to predict future outcomes or
outcomes for new subjects.

The function \code{predict()} allows us to obtain predicted values and
corresponding credible intervals from \code{JointAI} objects. Note that
for mixed models, currently only prediction for an ``average'' subject
is implemented, not prediction conditional on the random effects. A
dataset containing data for which the prediction should be performed is
specified via the argument \code{newdata}. If no \code{newdata} is
given, the original data are used. The argument \code{quantiles} allows
the specification of the quantiles of the posterior sample that are used
to obtain the credible interval (by default the 2.5\% and 97.5\%
quantile). Arguments \code{start}, \code{end}, \code{thin} and
\code{exclude_chains} control the subset of MCMC samples that is used.

\begin{CodeChunk}

\begin{CodeInput}
R> predict(mod13a, newdata = NHANES[27, ])
\end{CodeInput}

\begin{CodeOutput}
$newdata
         SBP gender age             race   WC alc educ creat albu
392 126.6667   male  32 Mexican American 94.1  <1  low  0.83  4.2
    uricacid bili occup  smoke      fit     2.5%    97.5%
392      8.7    1  <NA> former 116.3273 112.4343 120.1817

$fitted
       fit     2.5%    97.5%
1 116.3273 112.4343 120.1817
\end{CodeOutput}
\end{CodeChunk}

\code{predict()} returns a list with elements \code{newdata} (the
provided data with the predicted values and quantiles appended) and
\code{fit}, a \code{matrix} or \code{array} of the predicted values and
the quantiles that form the credible interval.

Via the argument \code{type} the user can specify the scale of the
predicted values. For generalized linear (mixed) models predicted values
can be calculated on the scale of the linear predictor
(\code{type = "link"} or \code{type =
"lp"}) or the scale of the response (\code{type = "response"}). For
ordinal and multinomial (mixed) models it is possible to return the
posterior probability of each of the outcome categories
(\code{type = "prob"}), the class with the highest mean posterior
probability (\code{type = "class"}, or \code{type =
"response"}) or the linear predictor (\code{type = "lp"}).

For parametric survival models \code{type = "lp"} is synonymous for
\code{type =
"link"} and \code{type = "linear"}, and \code{type = "response"} is
\code{exp(lp)}. The options for proportional hazards models are
\code{type =
"lp"}, \code{type = "risk"} (\(=\exp(\texttt{lp})\)),
\code{type = "survival"} and \code{type = "expected"}
(\(=-\log(\texttt{survival})\)).

\hypertarget{prediction-to-visualize-non-linear-effects}{%
\subsubsection{Prediction to visualize non-linear
effects}\label{prediction-to-visualize-non-linear-effects}}

Another reason to obtain predicted values is the visualization of
non-linear effects (see Figure~\ref{fig:predvis}). To facilitate the
generation of a dataset for such a prediction, the function
\code{predDF()} can be used. It generates a \code{data.frame} that
contains a sequence of values through the range of observed values for
the covariate specified by the argument \code{vars} which takes a
one-sided formula. Median or reference values are used for all other
continuous and categorical variables, respectively.

Create dataset for prediction and obtain predicted values:

\begin{CodeChunk}

\begin{CodeInput}
R> newDF <- predDF(mod13b, vars = ~age)
R> pred <- predict(mod13b, newdata = newDF)
\end{CodeInput}
\end{CodeChunk}

Plot predicted values and credible interval:

\begin{CodeChunk}

\begin{CodeInput}
R> matplot(pred$newdata$age, pred$newdata[, c("fit", "2.5%", "97.5%")],
+   lty = c(1,2,2), type = "l", col = 1, xlab = "age in months",
+   ylab = "predicted value")
\end{CodeInput}
\begin{figure}

{\centering \includegraphics[width=0.75\linewidth]{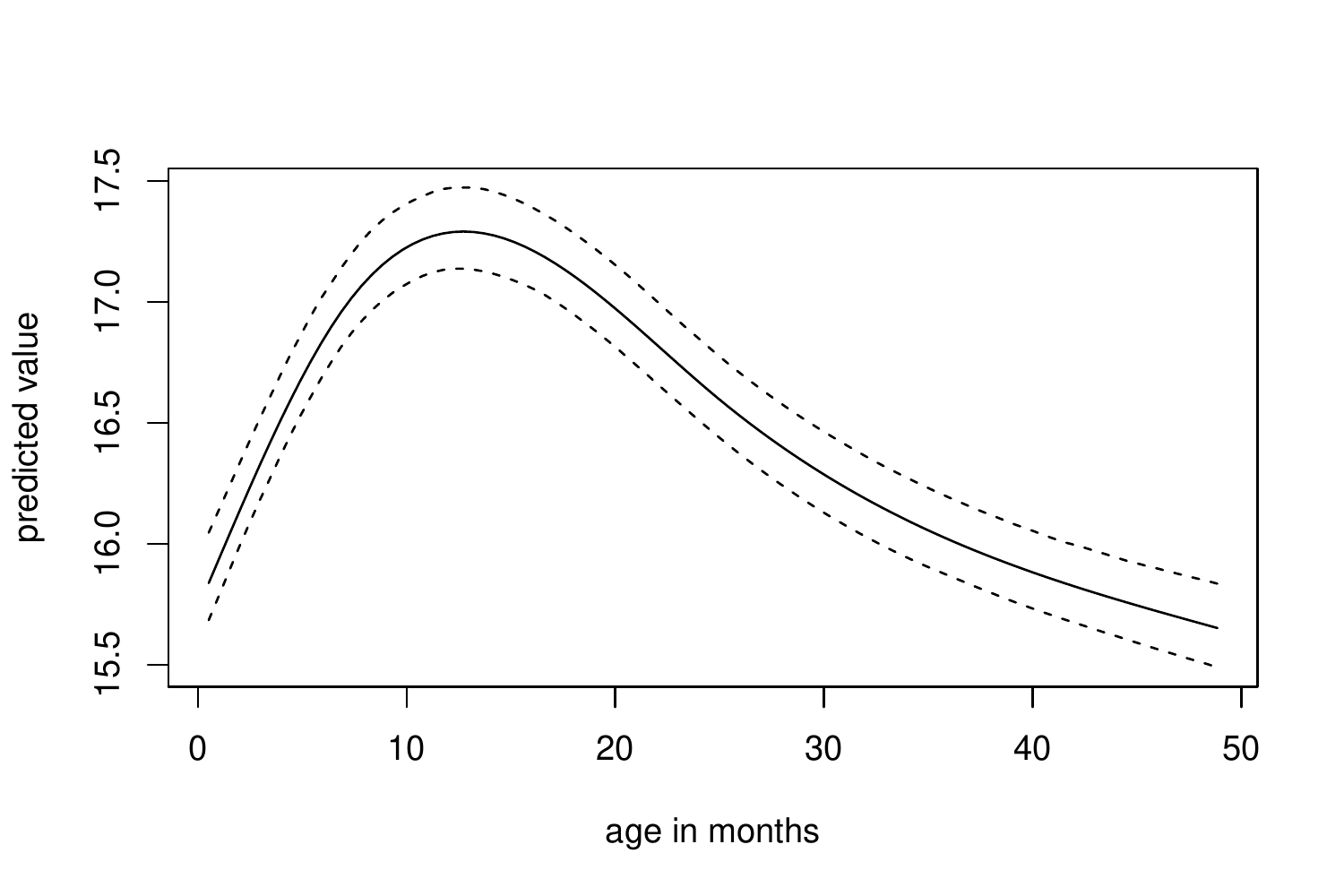} 

}

\caption[Predicted values of \code{BMI} and corresponding 95\% credible interval from \code{mod13b}]{Predicted values of \code{BMI} and corresponding 95\% credible interval from \code{mod13b}.}\label{fig:predvis}
\end{figure}
\end{CodeChunk}

The optional dots argument allows the user to explicitly specify which
values to be used for the variables given in \code{vars}, for example:

\begin{CodeChunk}

\begin{CodeInput}
R> newDF2 <- predDF(mod13b, vars = ~age + HEIGHT_M, HEIGHT_M = c(160, 175))
\end{CodeInput}
\end{CodeChunk}

\hypertarget{sec:getMIdat}{%
\subsection{Export of imputed values}\label{sec:getMIdat}}

Imputed datasets can be extracted from a \code{JointAI} object (in which
a monitor for the imputed values has been set, i.e.,
\code{monitor_params = c(imps
= TRUE)}) with the function \code{get_MIdat()}. It creates completed
datasets by taking the imputed values from randomly chosen iterations of
the MCMC sample and filling them into copies of the original incomplete
data.

The argument \code{m} specifies the number of imputed datasets to be
created, \code{include} controls whether the original data are included
in the long format \code{data.frame} (default is \code{include = TRUE}),
\code{start} specifies the first iteration that may be used, and
\code{minspace} is the minimum number of iterations between iterations
eligible for selection. To make the selection of iterations
reproducible, a seed value can be specified via the argument
\code{seed}.

When \code{export_to_SPSS = TRUE} the imputed data is exported to
\proglang{SPSS} (IBM SPSS Statistics, IBM Corp.), i.e., a \code{.txt}
file containing the data and a \code{.sps} file containing
\proglang{SPSS} syntax to convert the data into an SPSS data file (with
ending \code{.sav}) are written. Arguments \code{filename} and
\code{resdir} allow specification of the name of the \code{.txt} and
\code{.sps} file and the directory they are written to.

\code{get_MIdat()} returns a long-format \code{data.frame} containing
the imputed datasets (and by default the original data) stacked on top
of each other. The imputation number is given in the variable
\code{Imputation_}, column \code{.id} contains a newly created id
variable for each observation in cross-sectional data (multi-level data
should already contain an id variable) and the column \code{.rownr}
identifies rows of the original data (relevant in multi-level data).

\begin{CodeChunk}

\begin{CodeInput}
R> impDF <- get_MIdat(mod13d, m = 10, seed = 2019)
\end{CodeInput}
\end{CodeChunk}

The function \code{plot_imp_distr()} allows visual comparison of the
distributions of the observed and imputed values. The distribution of
the observed values is shown in dark blue, the distribution of the
imputed values per dataset in light blue (Figure~\ref{fig:impcomp}):

\begin{CodeChunk}

\begin{CodeInput}
R> plot_imp_distr(impDF, nrow = 1)
\end{CodeInput}
\begin{figure}

{\centering \includegraphics[width=1\linewidth]{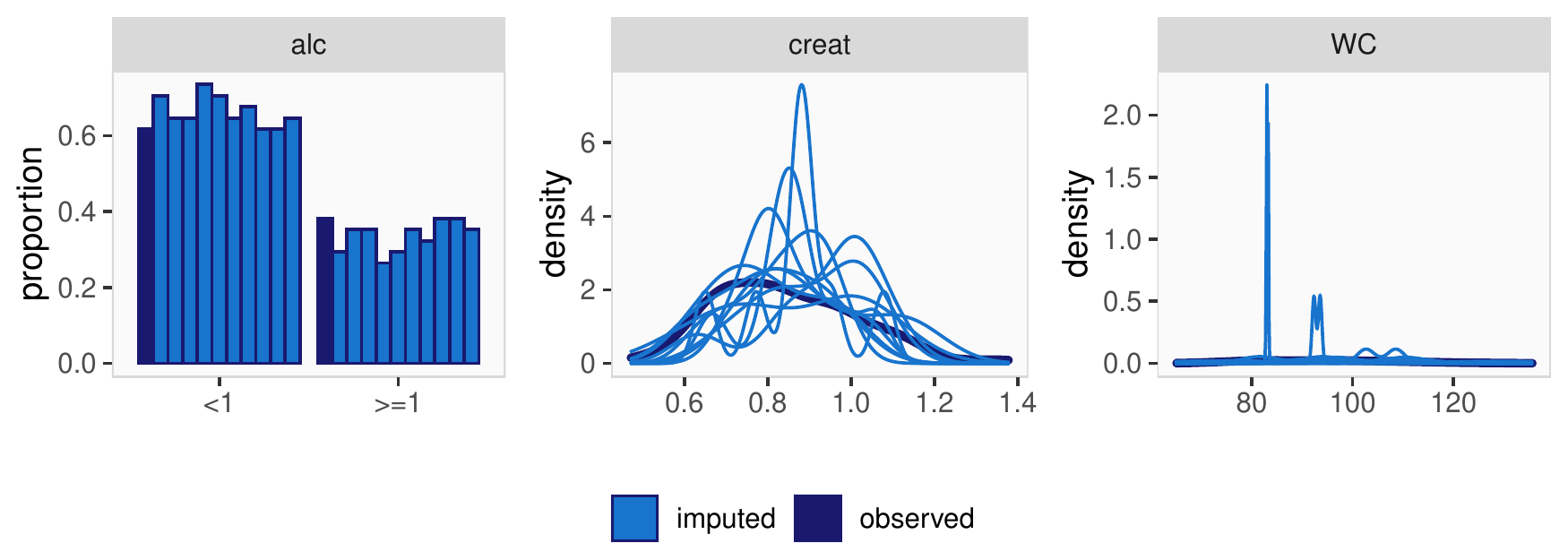} 

}

\caption[Distribution of observed and imputed values]{Distribution of observed and imputed values.}\label{fig:impcomp}
\end{figure}
\end{CodeChunk}

\pagebreak

\hypertarget{assumptions-and-extensions}{%
\section{Assumptions and extensions}\label{assumptions-and-extensions}}

Like any statistical model, the approach followed in \pkg{JointAI}
relies on assumptions that need to be satisfied in order to obtain valid
results.

A commonly made assumption that is also required for \pkg{JointAI} is
that the missing data mechanism is ignorable, i.e., that data is missing
at random (MAR) or missing completely at random (MCAR) \citep{Rubin1976}
and that parameters in the model of the missingness mechanism are
independent of the parameters in the data model \citep{Schafer1997}. It
is the task of the researcher to critically evaluate whether this
assumption is satisfied for a given dataset and model.

Furthermore, all models involved in the imputation and analysis need to
be correctly specified. In current implementations of imputation
procedures in software, imputation models are typically automatically
specified, using standard assumptions like linear associations and
default model types. In \pkg{JointAI}, the arguments \code{models} and
\code{auxvar} permit tailoring of the automatically chosen models to
some extent, by allowing the user to chose non-normal imputation models
for continuous variables and to include variables or functional forms of
variables that are not used in the analysis model in the linear
predictor of the imputation models. Moreover, it is possible to
explicitly specify the linear predictor of covariate models by providing
a list of model formulas instead of just the formula for the main
analysis model.

When using auxiliary variables in \pkg{JointAI}, it should be noted that
due to the default ordering of the conditional distributions in the
sequence of models it is implied that the auxiliary variable is
independent of the outcome, since neither the model for the auxiliary
variable has the outcome in its linear predictor nor vice versa. In some
settings it may be possible to avoid this assumption by providing a list
of model formulas in which the model for the auxiliary variable is
specified explicitly to include the outcome in its linear predictor.

Moreover, in order to make software usable, default values have to be
chosen for various parameters. These default values are chosen to work
well in certain settings, but cannot be guaranteed to be appropriate in
general and it is the task of the user to make the appropriate changes.
In \pkg{JointAI} this concerns, for example, the choice of
hyper-parameters and automatically chosen types of imputation models.

To expand the range of settings in which \pkg{JointAI} provides a valid
and user-friendly way to simultaneously analyse and impute data, several
extensions are planned. These include:

\begin{itemize}
\tightlist
\item
  Implementation of (penalized) splines for incompletely observed
  covariates.
\item
  Evaluation of model fit by providing functionality to perform
  posterior predictive checks.
\item
  Implementation of subject-specific prediction from mixed models.
\item
  Implementation of additional choices of shrinkage priors (such as
  lasso and elastic net).
\item
  Implementation of additional model types, for example, using
  zero-inflated or over-dispersed distributions.
\item
  Extensions of joint models for longitudinal and survival data to other
  association structures such as slopes and cumulative effects.
\item
  Implementation of different options for the random effects covariance
  matrix in multivariate mixed models.
\item
  Extensions of survival models to other types of censoring, competing
  risks and stratified baseline hazards.
\end{itemize}

\appendix

\hypertarget{sec:AppHyperpars}{%
\section{Default hyper-parameters}\label{sec:AppHyperpars}}

\begin{CodeChunk}

\begin{CodeInput}
R> default_hyperpars()
\end{CodeInput}

\begin{CodeOutput}
$norm
   mu_reg_norm   tau_reg_norm shape_tau_norm  rate_tau_norm 
         0e+00          1e-04          1e-02          1e-02 

$gamma
   mu_reg_gamma   tau_reg_gamma shape_tau_gamma  rate_tau_gamma 
          0e+00           1e-04           1e-02           1e-02 

$beta
   mu_reg_beta   tau_reg_beta shape_tau_beta  rate_tau_beta 
         0e+00          1e-04          1e-02          1e-02 

$binom
 mu_reg_binom tau_reg_binom 
        0e+00         1e-04 

$poisson
 mu_reg_poisson tau_reg_poisson 
          0e+00           1e-04 

$multinomial
 mu_reg_multinomial tau_reg_multinomial 
              0e+00               1e-04 

$ordinal
   mu_reg_ordinal   tau_reg_ordinal  mu_delta_ordinal 
            0e+00             1e-04             0e+00 
tau_delta_ordinal 
            1e-04 

$ranef
shape_diag_RinvD  rate_diag_RinvD       KinvD_expr 
          "0.01"          "0.001"   "nranef + 1.0" 

$surv
 mu_reg_surv tau_reg_surv 
       0.000        0.001 
\end{CodeOutput}
\end{CodeChunk}

\section[Density plot using ggplot2]{Density plot using \pkg{ggplot2}}\label{sec:AppDensplot}

Fit the complete-case version of the model:

\begin{CodeChunk}

\begin{CodeInput}
R> mod13a_cc <- lm(formula(mod13a), data = NHANES)
\end{CodeInput}
\end{CodeChunk}

Make a dataset containing the quantiles of the posterior sample and
confidence intervals from the complete case analysis:

\begin{CodeChunk}

\begin{CodeInput}
R> quantDF <- rbind(
+    data.frame(variable = rownames(summary(mod13a)$stat),
+               type = "2.5%",
+               model = "JointAI",
+               value = summary(mod13a)$stat[, c("2.5%")]),
+    data.frame(variable = rownames(summary(mod13a)$stat),
+               type = "97.5%",
+               model = "JointAI",
+               value = summary(mod13a)$stat[, c("97.5%")]),
+    data.frame(variable = names(coef(mod13a_cc)),
+               type = "2.5%",
+               model = "cc",
+               value = confint(mod13a_cc)[, "2.5 %"]),
+    data.frame(variable = names(coef(mod13a_cc)),
+               type = "97.5%",
+               model = "cc",
+               value = confint(mod13a_cc)[, "97.5 %"])
+  )
\end{CodeInput}
\end{CodeChunk}

\pkg{ggplot2} version, excluding \code{tau_SBP} from the plot:

\begin{CodeChunk}

\begin{CodeInput}
>R p13a <- densplot(mod13a, ncol = 3, use_ggplot = TRUE, joined = TRUE,
+    subset = c(analysis_main = TRUE, tau_y = FALSE)) +
+    theme(legend.position = "bottom")
\end{CodeInput}
\end{CodeChunk}

Add vertical lines for the:

\begin{itemize}
\tightlist
\item
  confidence intervals from the complete case analysis
\item
  quantiles of the posterior distribution
\end{itemize}

\begin{CodeChunk}

\begin{CodeInput}
R> p13a +
+    geom_vline(data = quantDF, aes(xintercept = value, color = model),
+    lty = 2) +
+    scale_color_manual(name = "CI from model: ", values = c("blue", "red"),
+                       limits = c("JointAI", "cc"),
+                       labels = c("JointAI", "compl.case"))
\end{CodeInput}
\end{CodeChunk}

\hypertarget{references}{%
\section{References}\label{references}}

\bibliography{JointAI.bib}

\begin{thebibliography}{44}
\newcommand{\enquote}[1]{``#1''}
\providecommand{\natexlab}[1]{#1}
\providecommand{\url}[1]{\texttt{#1}}
\providecommand{\urlprefix}{URL }
\expandafter\ifx\csname urlstyle\endcsname\relax
  \providecommand{\doi}[1]{doi:\discretionary{}{}{}#1}\else
  \providecommand{\doi}{doi:\discretionary{}{}{}\begingroup
  \urlstyle{rm}\Url}\fi
\providecommand{\eprint}[2][]{\url{#2}}

\bibitem[{Audigier and Resche-Rigon(2019)}]{micemd}
Audigier V, Resche-Rigon M (2019).
\newblock \emph{\pkg{micemd}: Multiple Imputation by Chained Equations with
  Multilevel Data}.
\newblock \proglang{R}~package version~1.6.0,
  \urlprefix\url{https://CRAN.R-project.org/package=micemd}.

\bibitem[{Bartlett and Keogh(2020)}]{smcfcs}
Bartlett J, Keogh R (2020).
\newblock \emph{\pkg{smcfcs}: Multiple Imputation of Covariates by Substantive
  Model Compatible Fully Conditional Specification}.
\newblock \proglang{R}~package version~1.4.1,
  \urlprefix\url{https://CRAN.R-project.org/package=smcfcs}.

\bibitem[{Bartlett \emph{et~al.}(2015)Bartlett, Seaman, White, and
  Carpenter}]{Bartlett2015}
Bartlett JW, Seaman SR, White IR, Carpenter JR (2015).
\newblock \enquote{{Multiple Imputation of Covariates by Fully Conditional
  Specification: Accommodating the Substantive Model}.}
\newblock \emph{Statistical Methods in Medical Research}, \textbf{24}(4),
  462--487.
\newblock \doi{10.1177/0962280214521348}.

\bibitem[{Bates \emph{et~al.}(2015)Bates, Machler, Bolker, and Walker}]{lme4}
Bates D, Machler M, Bolker B, Walker S (2015).
\newblock \enquote{Fitting Linear Mixed-Effects Models Using {lme4}.}
\newblock \emph{Journal of Statistical Software}, \textbf{67}(1), 1--48.
\newblock \doi{10.18637/jss.v067.i01}.

\bibitem[{Bengtsson(2020{\natexlab{a}})}]{doFuture}
Bengtsson H (2020{\natexlab{a}}).
\newblock \emph{doFuture: A Universal Foreach Parallel Adapter using the Future
  API of the 'future' Package}.
\newblock R package version 0.9.0,
  \urlprefix\url{https://CRAN.R-project.org/package=doFuture}.

\bibitem[{Bengtsson(2020{\natexlab{b}})}]{future}
Bengtsson H (2020{\natexlab{b}}).
\newblock \emph{future: Unified Parallel and Distributed Processing in R for
  Everyone}.
\newblock R package version 1.18.0,
  \urlprefix\url{https://CRAN.R-project.org/package=future}.

\bibitem[{Brooks and Gelman(1998)}]{Brooks1998}
Brooks SP, Gelman A (1998).
\newblock \enquote{{General Methods for Monitoring Convergence of Iterative
  Simulations}.}
\newblock \emph{Journal of Computational and Graphical Statistics},
  \textbf{7}(4), 434--455.
\newblock \doi{10.1080/10618600.1998.10474787}.

\bibitem[{Deng \emph{et~al.}(2016)Deng, Chang, Ido, and Long}]{Deng2016}
Deng Y, Chang C, Ido MS, Long Q (2016).
\newblock \enquote{{Multiple Imputation for General Missing Data Patterns in
  the Presence of High-dimensional Data}.}
\newblock \emph{Scientific reports}, \textbf{6}(1), 1--10.
\newblock \doi{10.1038/srep21689}.

\bibitem[{Erler(2020)}]{JointAI}
Erler NS (2020).
\newblock \emph{\pkg{JointAI}: Joint Analysis and Imputation of Incomplete
  Data}.
\newblock \proglang{R}~package version~1.0.0,
  \urlprefix\url{https://CRAN.R-project.org/package=JointAI}.

\bibitem[{Erler \emph{et~al.}(2019)Erler, Rizopoulos, Jaddoe, Franco, and
  Lesaffre}]{Erler2019}
Erler NS, Rizopoulos D, Jaddoe VW, Franco OH, Lesaffre EM (2019).
\newblock \enquote{{Bayesian Imputation of Time-Varying Covariates in Linear
  Mixed Models}.}
\newblock \emph{Statistical Methods in Medical Research}, \textbf{28}(2), 555
  -- 568.
\newblock \doi{10.1177/0962280217730851}.

\bibitem[{Erler \emph{et~al.}(2016)Erler, Rizopoulos, van Rosmalen, Jaddoe,
  Franco, and Lesaffre}]{Erler2016}
Erler NS, Rizopoulos D, van Rosmalen J, Jaddoe VW, Franco OH, Lesaffre EM
  (2016).
\newblock \enquote{{Dealing with Missing Covariates in Epidemiologic Studies: A
  Comparison between Multiple Imputation and a Full Bayesian Approach}.}
\newblock \emph{Statistics in Medicine}, \textbf{35}(17), 2955--2974.
\newblock \doi{10.1002/sim.6944}.

\bibitem[{Flegal \emph{et~al.}(2020)Flegal, Hughes, Vats, and Dai}]{mcmcse}
Flegal JM, Hughes J, Vats D, Dai N (2020).
\newblock \emph{\pkg{mcmcse}: Monte Carlo Standard Errors for MCMC}.
\newblock \proglang{R}~package version~1.4-1,
  \urlprefix\url{https://CRAN.R-project.org/package=mcmcse}.

\bibitem[{Gelman and Rubin(1992)}]{Gelman1992}
Gelman A, Rubin DB (1992).
\newblock \enquote{{Inference from Iterative Simulation Using Multiple
  Sequences}.}
\newblock \emph{Statistical Science}, \textbf{7}(4), 457--472.
\newblock \doi{10.1214/ss/1177011136}.

\bibitem[{Geraci and McLain(2018)}]{Geraci2018}
Geraci M, McLain A (2018).
\newblock \enquote{{Multiple Imputation for Bounded Variables}.}
\newblock \emph{Psychometrika}, \textbf{83}(4), 919--940.
\newblock \doi{10.1007/s11336-018-9616-y}.

\bibitem[{Grund \emph{et~al.}(2019)Grund, Robitzsch, and Lüdtke}]{mitml}
Grund S, Robitzsch A, Lüdtke O (2019).
\newblock \emph{\pkg{mitml}: Tools for Multiple Imputation in Multilevel
  Modeling}.
\newblock \proglang{R}~package version~0.3-7,
  \urlprefix\url{https://CRAN.R-project.org/package=mitml}.

\bibitem[{Hadfield(2010)}]{MCMCglmm}
Hadfield JD (2010).
\newblock \enquote{MCMC Methods for Multi-Response Generalized Linear Mixed
  Models: The \pkg{MCMCglmm} \proglang{R} Package.}
\newblock \emph{Journal of Statistical Software}, \textbf{33}(2), 1--22.
\newblock \urlprefix\url{http://www.jstatsoft.org/v33/i02/}.

\bibitem[{Ibrahim \emph{et~al.}(2002)Ibrahim, Chen, and Lipsitz}]{Ibrahim2002}
Ibrahim JG, Chen MH, Lipsitz SR (2002).
\newblock \enquote{Bayesian Methods for Generalized Linear Models with
  Covariates Missing At Random.}
\newblock \emph{Canadian Journal of Statistics}, \textbf{30}(1), 55--78.
\newblock \doi{10.2307/3315865}.

\bibitem[{Kowarik and Templ(2016)}]{VIM}
Kowarik A, Templ M (2016).
\newblock \enquote{Imputation with the \proglang{R} Package \pkg{VIM}.}
\newblock \emph{Journal of Statistical Software}, \textbf{74}(7), 1--16.
\newblock \doi{10.18637/jss.v074.i07}.
\newblock \proglang{R}~package version~4.8.0.

\bibitem[{Lesaffre and Lawson(2012)}]{Lesaffre2012}
Lesaffre E, Lawson AB (2012).
\newblock \emph{Bayesian Biostatistics}.
\newblock John Wiley \& Sons.
\newblock ISBN 978-0-470-01823-1.
\newblock \doi{10.1002/9781119942412}.

\bibitem[{{National Center for Health Statistics (NCHS)}(2011 --
  2012)}]{NHANES2011}
{National Center for Health Statistics (NCHS)} (2011 -- 2012).
\newblock \enquote{National Health and Nutrition Examination Survey Data.}
\newblock \urlprefix\url{https://www.cdc.gov/nchs/nhanes/}.

\bibitem[{Novo and Schafer(2013)}]{norm}
Novo AA, Schafer JL (2013).
\newblock \emph{\pkg{norm}: Analysis of Multivariate Normal Datasets with
  Missing Values}.
\newblock \proglang{R}~package version~1.0.9.5,
  \urlprefix\url{https://CRAN.R-project.org/package=norm}.

\bibitem[{Pinheiro \emph{et~al.}(2020)Pinheiro, Bates, DebRoy, Sarkar, and {R
  Core Team}}]{nlme}
Pinheiro J, Bates D, DebRoy S, Sarkar D, {R Core Team} (2020).
\newblock \emph{\pkg{nlme}: Linear and Nonlinear Mixed Effects Models}.
\newblock \proglang{R}~package version 3.1.148,
  \urlprefix\url{https://CRAN.R-project.org/package=nlme}.

\bibitem[{Plummer(2003)}]{JAGS}
Plummer M (2003).
\newblock \enquote{{\proglang{JAGS}: A Program for Analysis of Bayesian
  Graphical Models using Gibbs Sampling}.}
\newblock In K~Hornik, F~Leisch, A~Zeileis (eds.), \emph{Proceedings of the 3rd
  International Workshop on Distributed Statistical Computing (DSC 2003)}.
\newblock {ISSN}: 1609-395X.

\bibitem[{Plummer(2017)}]{JAGSmanual}
Plummer M (2017).
\newblock \emph{{\proglang{JAGS} Version~4.3.0 User Manual}}.
\newblock
  \urlprefix\url{https://sourceforge.net/projects/mcmc-jags/files/Manuals/4.x/jags_user_manual.pdf/download}.

\bibitem[{Plummer(2019)}]{rjags}
Plummer M (2019).
\newblock \emph{\pkg{rjags}: Bayesian Graphical Models using MCMC}.
\newblock \proglang{R}~package version 4.10,
  \urlprefix\url{https://CRAN.R-project.org/package=rjags}.

\bibitem[{Plummer \emph{et~al.}(2006)Plummer, Best, Cowles, and Vines}]{coda}
Plummer M, Best N, Cowles K, Vines K (2006).
\newblock \enquote{CODA: Convergence Diagnosis and Output Analysis for MCMC.}
\newblock \emph{\proglang{R} News}, \textbf{6}(1), 7--11.
\newblock \urlprefix\url{https://journal.r-project.org/archive/}.

\bibitem[{Quartagno and Carpenter(2020)}]{jomo}
Quartagno M, Carpenter J (2020).
\newblock \emph{\pkg{jomo}: A Package for Multilevel Joint Modelling Multiple
  Imputation}.
\newblock \proglang{R}~package version~2.7.2,
  \urlprefix\url{https://CRAN.R-project.org/package=jomo}.

\bibitem[{{\proglang{R} Core Team}(2020)}]{RVersion}
{\proglang{R} Core Team} (2020).
\newblock \emph{\proglang{R}: A Language and Environment for Statistical
  Computing}.
\newblock \proglang{R} Foundation for Statistical Computing, Vienna, Austria.
\newblock \urlprefix\url{https://www.R-project.org/}.

\bibitem[{Robitzsch \emph{et~al.}(2020)Robitzsch, Grund, and Henke}]{miceadds}
Robitzsch A, Grund S, Henke T (2020).
\newblock \emph{\pkg{miceadds}: Some Additional Multiple Imputation Functions,
  Especially for \pkg{mice}}.
\newblock \proglang{R}~package version~3.10-28,
  \urlprefix\url{https://CRAN.R-project.org/package=miceadds}.

\bibitem[{Robitzsch and Lüdtke(2020)}]{mdmb}
Robitzsch A, Lüdtke O (2020).
\newblock \emph{\pkg{mdmb}: Model Based Treatment of Missing Data}.
\newblock \proglang{R}~package version~1.4.12,
  \urlprefix\url{https://CRAN.R-project.org/package=mdmb}.

\bibitem[{Rodwell \emph{et~al.}(2014)Rodwell, Lee, Romaniuk, and
  Carlin}]{Rodwell2014}
Rodwell L, Lee KJ, Romaniuk H, Carlin JB (2014).
\newblock \enquote{{Comparison of Methods for Imputing Limited-range Variables:
  A Simulation Study}.}
\newblock \emph{BMC Medical Research Methodology}, \textbf{14}(1), 57.
\newblock \doi{10.1186/1471-2288-14-57}.

\bibitem[{Rubin(1987)}]{Rubin1987}
Rubin D (1987).
\newblock \emph{{Multiple Imputation for Nonresponse in Surveys}}.
\newblock John Wiley \& Sons.

\bibitem[{Rubin(1976)}]{Rubin1976}
Rubin DB (1976).
\newblock \enquote{{Inference and Missing Data}.}
\newblock \emph{Biometrika}, \textbf{63}(3), 581--592.
\newblock \doi{10.2307/2335739}.

\bibitem[{Rubin(2004)}]{Rubin2004}
Rubin DB (2004).
\newblock \enquote{The {D}esign of a {G}eneral and {F}lexible {S}ystem for
  {H}andling {N}onresponse in {S}ample {S}urveys.}
\newblock \emph{The American Statistician}, \textbf{58}, 298--302.
\newblock \doi{10.1198/000313004X6355}.

\bibitem[{Schafer(1997)}]{Schafer1997}
Schafer JL (1997).
\newblock \emph{{Analysis of Incomplete Multivariate Data}}.
\newblock Chapman \& Hall/CRC, New York.

\bibitem[{Speidel \emph{et~al.}(2020)Speidel, Drechsler, and Jolani}]{hmi}
Speidel M, Drechsler J, Jolani S (2020).
\newblock \emph{\pkg{hmi}: Hierarchical Multiple Imputation}.
\newblock \proglang{R}~package version~0.9.19,
  \urlprefix\url{https://CRAN.R-project.org/package=hmi}.

\bibitem[{Tierney \emph{et~al.}(2020)Tierney, Cook, McBain, and Fay}]{naniar}
Tierney N, Cook D, McBain M, Fay C (2020).
\newblock \emph{\pkg{naniar}: Data Structures, Summaries, and Visualisations
  for Missing Data}.
\newblock \proglang{R}~package version~0.5.2,
  \urlprefix\url{https://CRAN.R-project.org/package=naniar}.

\bibitem[{Treiman(2009)}]{Treiman2009}
Treiman D (2009).
\newblock \emph{{Quantitative Data Analysis: Doing Social Research to Test
  Ideas}}.
\newblock Research Methods for the Social Sciences. Wiley.
\newblock ISBN 9780470380031.
\newblock \urlprefix\url{https://books.google.com.vc/books?id=ybYMUgCHzC0C}.

\bibitem[{Van~Buuren(2012)}]{Buuren2012}
Van~Buuren S (2012).
\newblock \emph{{Flexible Imputation of Missing Data}}.
\newblock Taylor \& Francis.

\bibitem[{{Van Buuren} and Groothuis-Oudshoorn(2011)}]{mice}
{Van Buuren} S, Groothuis-Oudshoorn K (2011).
\newblock \enquote{\pkg{mice}: Multivariate Imputation by Chained Equations in
  \proglang{R}.}
\newblock \emph{Journal of Statistical Software}, \textbf{45}(3), 1--67.
\newblock \urlprefix\url{https://www.jstatsoft.org/v45/i03/}.

\bibitem[{Von~Hippel(2013)}]{vonHippel2012}
Von~Hippel PT (2013).
\newblock \enquote{{Should a Normal Imputation Model be Modified to Impute
  Skewed Variables?}}
\newblock \emph{Sociological Methods \& Research}, \textbf{42}(1), 105--138.
\newblock \doi{10.1177/0049124112464866}.

\bibitem[{White \emph{et~al.}(2011)White, Royston, and Wood}]{White2011}
White IR, Royston P, Wood AM (2011).
\newblock \enquote{Multiple Imputation using Chained Equations: Issues and
  Guidance for Practice.}
\newblock \emph{Statistics in Medicine}, \textbf{30}(4), 377--399.
\newblock \doi{10.1002/sim.4067}.

\bibitem[{Wickham(2016)}]{ggplot2}
Wickham H (2016).
\newblock \emph{\pkg{ggplot2}: Elegant Graphics for Data Analysis}.
\newblock Springer-Verlag, New York.
\newblock ISBN 978-3-319-24277-4.
\newblock \urlprefix\url{https://ggplot2.tidyverse.org}.

\bibitem[{Yucel(2010)}]{mlmmm}
Yucel R (2010).
\newblock \emph{\pkg{mlmmm}: ML Estimation under Multivariate Linear Mixed
  Models with Missing Values}.
\newblock \proglang{R}~package version~0.3.1.2,
  \urlprefix\url{https://CRAN.R-project.org/package=mlmmm}.

\end{thebibliography}

\end{document}